\documentclass[acmsmall,screen]{acmart}
\newcommand{\rev}[1]{\textcolor{black}{#1}}
\usepackage{tikz}
\usepackage{pgfplots}
\usetikzlibrary{patterns}
\pgfplotsset{compat=1.18}
\usepackage{amsmath,amsfonts}
\usepackage{tikz}
\usepackage{algorithm}
\usepackage[caption=false,font=normalsize,labelfont=sf,textfont=sf]{subfig}
\usepackage{textcomp}
\usepackage{stfloats}
\usepackage{url}
\usepackage{verbatim}
\usepackage{graphicx}
\usepackage{booktabs} 
\usepackage{pifont}
\usepackage{caption}
\usepackage{algpseudocode}
\usepackage{multirow}
\usetikzlibrary{calc}
\usepackage{adjustbox}
\usepackage{array}
\usepackage{balance}
\usepackage{amsmath,amsfonts}
\usepackage{xcolor}
\usepackage{textcomp}
\usepgfplotslibrary{fillbetween}
\usepackage{xcolor}
\usepackage{tabularx}

\newcolumntype{L}{>{\raggedright\arraybackslash}X}
\newcolumntype{C}{>{\centering\arraybackslash}X}

\AtBeginDocument{%
  }

\setcopyright{acmlicensed}
\copyrightyear{2018}
\acmYear{2026}
\acmDOI{XXXXXXX.XXXXXXX}

\acmJournal{JACM}
\acmVolume{37}
\acmNumber{4}
\acmArticle{111}
\acmMonth{8}

\begin{document}

\title{Hardware-Rooted PUF Fingerprinting for Device-Level Traceability in Knowledge Distillation}

\author{Ning Lyu}
\email{nil418@lehigh.edu}
\orcid{0000-0001-9102-6879}
\affiliation{%
  \institution{Lehigh University}
  \streetaddress{27 Memorial Dr W}
  \city{Bethlehem}
  \state{PA}
  \country{USA}
  \postcode{18015}
}

\author{Yuntao Liu}
\email{yule24@lehigh.edu}
\orcid{0000-0001-8213-582X}
\affiliation{%
  \institution{Lehigh University}
  \streetaddress{27 Memorial Dr W}
  \city{Bethlehem}
  \state{PA}
  \country{USA}}

\author{Yonghong Bai}
\affiliation{%
  \institution{Broadcom Inc.}
  \streetaddress{Broadcom Inc}
  \city{Palo Alto}
  \state{CA}
  \country{USA}
}

\author{Zhiyuan Yan}
\affiliation{%
  \institution{Lehigh University}
  \streetaddress{27 Memorial Dr W}
  \city{Bethlehem}
  \state{PA}
  \country{USA}}

\renewcommand{\shortauthors}{Ning Lyu et al.}

\begin{abstract}
  Knowledge distillation (KD) enables model transfer across heterogeneous platforms and deployment environments, yet it exposes proprietary models to distillation-based theft, where an adversary extracts intellectual property (IP) by training a student model on teacher outputs. Current defenses, such as software watermarking or hardware-based access control, either fail to survive the distillation process or lack the granularity to identify the specific device responsible for a leak. Aiming to provide post-theft accountability and traceability, we propose a novel fingerprinting framework that superimposes device-specific Physical Unclonable Function (PUF) signatures onto teacher logits during distillation. By utilizing signatures derived from Ring Oscillator (RO) PUFs measured on a Xilinx Zynq-7020 FPGA, we ensure that any student model trained via KD inherits a unique, hardware-linked identity. Our framework is architecture-agnostic, enabling reliable identity inheritance across heterogeneous structures, including Convolutional Neural Networks, Vision Transformers, and encoders. To ensure robust attribution, we implement a two-stage recovery pipeline consisting of a neural decoder and Hamming-distance refinement, maintaining high detection accuracy even under noisy conditions. Furthermore, we introduce a multi-level logit encoding scheme to support large-scale device deployment. Experimental results demonstrate that the embedded fingerprints are resilient against common post-distillation modifications. These results establish a practical system-level approach for enabling hardware-linked model traceability in distributed AI deployment environments.
\end{abstract}



\maketitle

\section{Introduction}
\label{sec:introduction}

Knowledge distillation (KD) has become a fundamental technique in modern machine learning, enabling the effective transfer of predictive power from high-capacity teacher models to student models\cite{distillingknowle}. By leveraging the teacher's soft labels, student models can achieve competitive generalization performance with significantly reduced computational and memory requirements. This paradigm is particularly vital for edge deployment on platforms such as Field Programmable Gate Arrays (FPGAs), IoT systems, and embedded hardware, where teachers are increasingly deployed locally to facilitate real-time fine-tuning, continual learning, and adaptive AI applications without persistent cloud reliance.

However, deploying proprietary teacher models on edge hardware introduces a significant security risk. Once a model is mapped onto hardware, its intellectual property (IP), which represents immense training and architectural investment, is exposed to adversaries. 
Beyond traditional model extraction\cite{learningmodels, reverse}, distillation-based theft presents a sophisticated threat: an attacker can probe the teacher, collect input-output response pairs, and train a surrogate student model to replicate its behavior. This distillation-based theft enables unauthorized reproduction of proprietary model functionality.

Prior work has explored digital watermarking and ownership protection. Recent approaches improve robustness against functionality-stealing attacks, including model extraction and distillation, through margin-based watermarking and transferable trigger-set design\cite{margin,trigger_set}. Nevertheless, these approaches remain software-level protections: although effective for model ownership verification, they do not inherently bind the learned behavior to a unique physical device, and therefore offer limited support for device-level attribution after distillation-based leakage. In addition, distillation-aware defenses such as Safe Distillation Box seek to prevent unauthorized KD, but do not address post-leakage hardware-rooted traceability\cite{safe_kd}.

Recent hardware-oriented defenses have emerged to secure deep learning systems. Grailoo \textit{et al}. \cite{grailoo2022} introduced prediction poisoning via truncation-based noise to obstruct unauthorized distillation; however, this approach inadvertently restricts legitimate local optimization for edge devices. Olney and Karam \cite{olney2022protecting} proposed hardware-locked input obfuscation, yet this mechanism offers no protection once the model is unlocked by an authorized user. Furthermore, while DeepHardMark \cite{deephardmark} employs accelerator-level watermarking for ownership verification, it lacks the device-specific traceability required to identify the source of leaked distilled models. \rev{A detailed comparison with representative software- and hardware-based
model-protection approaches is provided in Table~\ref{tab:prior_comparison},
highlighting differences in protection objective, verification access,
hardware awareness, and support for device-level attribution.}


\rev{Consequently, existing software and hardware protections may verify model theft or restrict unauthorized access, but they generally cannot identify which authorized deployment was the source of a leaked distilled model. This is important when the same proprietary teacher is deployed across many hardware instances. If a suspicious student model can be linked to one enrolled device identity, the owner can narrow the investigation to that deployment and take targeted actions such as revoking access, rotating credentials, or inspecting the affected environment. Our framework provides this capability by embedding a hardware-rooted signature into the teacher outputs, allowing the distilled student to inherit a device-specific fingerprint for post-leak attribution. }

While previous works \cite{li2020puf, ISCAS} have utilized PUFs for model binding or hardware authentication, they primarily address runtime access control. To support post-leak device-level attribution, we introduce a lightweight fingerprinting framework that superimposes PUF signatures onto teacher logits via structured perturbations. \rev{Our method embeds the hardware identity at the model output level, allowing the fingerprint to be transferred and recovered even when the teacher and student use different architectures.}

The main contributions of this work are summarized as follows:
\begin{itemize}

\item \textbf{Hardware-Rooted Logit Fingerprinting:} 
We introduce a fingerprinting framework that embeds device identities derived from physical PUF measurements on a Xilinx Zynq-7020 (ZedBoard) FPGA directly into the teacher model logits. By incorporating real hardware measurements into the training process, the method enables device-level fingerprinting while capturing non-ideal behaviors of physical PUFs, including process variation, routing asymmetry, and environmental effects.

\item \rev{\textbf{Cross-Architecture Fingerprint Inheritance:} 
 We demonstrate that the embedded fin-
gerprint can propagate through knowledge distillation across heterogeneous architectures,
including Encoder-to-Encoder, Transformer-to-CNN, and CNN-to-CNN settings. Conse-
quently, the embedded hardware identity can be consistently recovered from the student
model, even under cross-architecture distillation.}

\item \textbf{Robust Two-Stage Identity Recovery:} 
We develop a two-stage recovery pipeline consisting of a Stage-I neural decoder for bit prediction and a Stage-II Hamming-distance refinement module for error correction. To eliminate the need for real distilled student models from every enrolled device, we introduce a synthetic logit simulation strategy to emulate the statistical effects of fingerprint embedding and distillation noise, enabling scalable decoder training. This hierarchical recovery framework significantly improves robustness and supports accurate device attribution under noisy conditions.

\item \textbf{Scalable PUF-to-Logit Encoding:}
We introduce a scalable PUF-to-logit encoding framework that adapts to different relationships between the hardware signature length and the model output dimension. When the two dimensions match, the signature is directly embedded into the logits. When extra logit dimensions are available, the signature is distributed across the larger logit space to improve robustness. When the desired hardware identity exceeds the output dimension, multi-level encoding packs multiple PUF bits into each output value for high-capacity attribution.

\item \textbf{Robustness to Post-Distillation Modifications:} 
We evaluate whether the embedded signatures remain recoverable under common post-distillation changes, including in-domain fine-tuning, cross-domain adaptation, and latent-space perturbations.
Our results show that the embedded identity is stable under standard fine-tuning, and can only be weakened through either extended retraining or direct perturbations to the latent space. In practice, these modifications either require significant additional training effort or lead to a noticeable drop in model performance.

\end{itemize}

The remainder of this paper is organized as follows: Section II discusses background in hardware security and model protection. Section III details the proposed PUF-Logit framework. Section IV presents comprehensive experimental results. Finally, Section V concludes the paper.

\section{Background and Preliminaries}
\label{sec:background}

\subsection{Knowledge Distillation}
Knowledge Distillation, originally formalized by Hinton \textit{et al.}~\cite{distillingknowle}, has evolved into a foundational paradigm for model compression and efficient deployment. In modern AI ecosystems, high-performance teacher models are often implemented on hardware, such as FPGAs. KD enables a compact student model to inherit the predictive power of these hardware-accelerated teachers by leveraging the latent information embedded in the teacher's softened output distributions.

Unlike standard supervised learning that relies on hard categorical labels, KD utilizes the \textit{logits} ($\mathbf{z}$), which are the raw, unnormalized outputs of the final fully connected layer. These logits capture the teacher's perception of inter-class similarities. To extract this information, the teacher's logit vector $\mathbf{z}_t$ and the student's logit vector $\mathbf{z}_s$ are transformed using a \textit{temperature-scaled softmax} function. This produces softened probability distributions $\mathbf{p}_t$ and $\mathbf{p}_s$, defined as:
\begin{equation}
\mathbf{p}_t = \text{Softmax}\left(\frac{\mathbf{z}_t}{T}\right), \quad \mathbf{p}_s = \text{Softmax}\left(\frac{\mathbf{z}_s}{T}\right)
\end{equation}
where $T$ is the temperature hyperparameter. As $T$ increases, the resulting distribution becomes more uniform, effectively amplifying the signals from the non-target classes. The student model is trained to minimize the information discrepancy between these distributions, typically measured via the \textbf{Kullback–Leibler (KL) divergence}:
\begin{equation}
\label{eq:kl_divergence}
\mathcal{L}_{\text{KD}} = T^2 \cdot \sum_{i=1}^{d} p_{t,i} \log \left( \frac{p_{t,i}}{p_{s,i}} \right)
\end{equation}
where $d$ denotes the number of classes in the label space. 

Although knowledge distillation is used for efficient model transfer, it can also serve as a potential vector for intellectual property theft. By querying a deployed teacher model and using its output responses as supervision, an adversary may train an unauthorized student model that approximates the teacher’s behavior. Consequently, the functional knowledge of the proprietary model can be replicated without exposing its internal parameters or architecture.

\subsection{PUFs as Hardware Fingerprints}
\label{sec:puf_background}

A PUF is a hardware security component that exploits the inherent, stochastic physical variations introduced during the integrated circuit (IC) manufacturing process.
These microscopic deviations, such as gate-level delay variations, threshold voltage shifts, or interconnect impedance, are unpredictable and uncontrollable, even by the manufacturer. Consequently, a PUF serves as a ``silicon fingerprint" unique to an individual hardware instance.

In FPGA-based deployments, various architectures are utilized to extract this entropy. \textit{SRAM-based PUFs} ~\cite{sram1, FPGA, bai2025new} leverage the random power-up states of memory cells~\cite{sram}, while \textit{Ring Oscillator (RO) PUFs} compare the frequency differences between multiple oscillating loops~\cite{ro}. Alternatively, \textit{Arbiter PUFs} exploit timing differences in parallel delay paths~\cite{arb, suh2007physical, liu2016optimization}. Unlike traditional non-volatile memory (NVM), PUFs generate cryptographic keys \textit{on-demand} only when the circuit is powered, significantly reducing the risk of physical probing and permanent key leakage.

While PUFs provide device-unique identities derived from manufacturing variations in silicon, their responses are not perfectly stable under real conditions. In practical deployments, two hardware characteristics must be considered when using PUF responses for device-level fingerprinting: \textbf{intra-device noise} and \textbf{reduced effective entropy}. 

\textbf{1). PUF Reliability and Uniqueness:}  
Environmental factors such as temperature variation, supply voltage fluctuations, and silicon aging can introduce bit flips in PUF responses. Prior studies report bit-error rates of approximately $1\%$--$5\%$~\cite{PUF_NOISE, variation, variation1, bai2017physical}. To ensure reliable device attribution, our framework incorporates noise characteristics measured from the FPGA RO-PUF during training, enabling robust fingerprint recovery under realistic hardware conditions. \rev{Different FPGA instances, even within the same device family, are expected to produce different PUF responses because their signatures arise from instance-specific manufacturing variations. For a given device, voltage and temperature changes may cause a limited number of unstable bits to flip; these variations are treated as intra-device response noise rather than a change of identity.}

\textbf{2). Effective Key Space and Entropy:}  
Although an \(n\)-bit PUF theoretically supports \(2^n\) unique identifiers, the effective entropy may be lower due to bit bias, spatial correlations, and unstable bit positions~\cite{instable}. For example, a nominal 10-bit PUF may provide an effective entropy closer to 8 bits, corresponding to roughly \(2^8=256\) distinguishable identities. Such bias or correlation can also reduce the Hamming-distance separation between enrolled PUF identities, which may decrease the attribution margin during Stage-II matching. In practical enrollment, this effect can be mitigated by selecting stable and discriminative PUF bits and, when necessary, increasing the fingerprint length.

By accounting for both response noise and reduced entropy, our framework ensures that the embedded device signatures remain reliable and traceable in practical deployments.

\subsection{Representative Model Architectures}
\rev{To evaluate whether the embedded hardware fingerprint remains recoverable across different teacher–student structures, we consider three representative architectural settings.} By injecting device-specific signatures at the behavioral output (logits) rather than modifying internal layers, our framework maintains compatibility with heterogeneous backbones.

\textit{Convolutional Neural Networks (CNNs)}:
CNNs remain the standard for computer vision tasks due to their inductive bias of translation invariance and local feature extraction. In a typical \textit{CNN-to-CNN} distillation setup, both the teacher and student utilize convolutional layers to process spatial hierarchies. Our framework leverages the logit outputs of these networks, ensuring that the spatial feature maps remain undisturbed while the final classification behavior carries the hardware-rooted signature.

\textit{Vision Transformers (ViTs)} \cite{dosovitskiy2021an}:
Transformer-to-CNN distillation has become a common setting for cross-architecture knowledge transfer. Vision Transformers use self-attention mechanisms to capture global relationships in the input, often producing strong teacher signals for distillation. Evaluating this setting allows us to test whether the PUF-based fingerprint remains inheritable when the teacher and student use different internal architectures.

\textit{Representation Learning and Encoders} \cite{NEURIPS2020_975a1c8b}:
In many edge scenarios, the goal is not direct classification but rather generating robust embeddings for downstream tasks. \textit{Encoder-to-Encoder} distillation (often found in self-supervised or metric learning) focuses on the latent-space representation. Extending our framework to this setting demonstrates that the PUF signature can be superimposed onto continuous embedding spaces, ensuring traceability even when traditional class-based logits are absent.

\subsection{Comparison with Existing Model Protection Methods}

\rev{Table~\ref{tab:prior_comparison} summarizes representative software-
and hardware-based model-protection approaches. Since these methods were
developed under different threat models and protection objectives, we compare
their primary purpose, ownership or leakage detection capability, support for
black-box verification, use of hardware-specific information, and ability to
provide post-leak device-level attribution. The comparison highlights that the
proposed framework is complementary to conventional ownership-verification and
model-protection mechanisms by providing hardware-instance-level source
attribution after leakage.}

\begin{table}[t]
\centering
\caption{Comparison with representative model protection and ownership-verification approaches.}
\label{tab:prior_comparison}
\resizebox{\linewidth}{!}{%
\begin{tabular}{lccccc}
\hline
\textbf{Method} &
\textbf{\shortstack{Primary\\Objective}} &
\textbf{\shortstack{Ownership /\\Leakage Detection}} &
\textbf{\shortstack{Black-Box\\Verification}} &
\textbf{\shortstack{Hardware\\Aware}} &
\textbf{\shortstack{Device-Level\\Attribution}} \\
\hline

Adi et al.~\cite{USENIX} &
Watermarking &
$\checkmark$ &
$\checkmark$ &
$\times$ &
$\times$ \\

Uchida et al.~\cite{Uchida_2017} &
Watermarking &
$\checkmark$ &
$\times$ &
$\times$ &
$\times$ \\

Grailoo et al.~\cite{grailoo2022} &
KD theft prevention &
$\times$ &
-- &
$\checkmark$ &
$\times$ \\

Olney and Karam~\cite{olney2022protecting} &
Model IP protection &
$\times$ &
-- &
$\checkmark$ &
$\times$ \\

Jiang et al.~\cite{ISCAS} &
PUF-based IP protection &
$\checkmark$ &
$\checkmark$ &
$\checkmark$ &
$\times$ \\

Clements and Lao~\cite{deephardmark} &
Hardware watermarking &
$\checkmark$ &
-- &
$\checkmark$ &
$\times$ \\

\textbf{PUF-Logit (Ours)} &
\textbf{Source attribution} &
\textbf{$\checkmark$} &
\textbf{$\checkmark$} &
\textbf{$\checkmark$} &
\textbf{$\checkmark$} \\

\hline
\end{tabular}%
}
\vspace{1mm}
\begin{flushleft}
\footnotesize
$\checkmark$ indicates that the capability is supported by the method's
intended design; $\times$ indicates that it is not provided; ``--'' indicates
that the criterion is not directly applicable to the method's primary
protection objective.
\end{flushleft}
\end{table}

\begin{figure}[t]
\centering
\includegraphics[width=\linewidth,height=7cm]{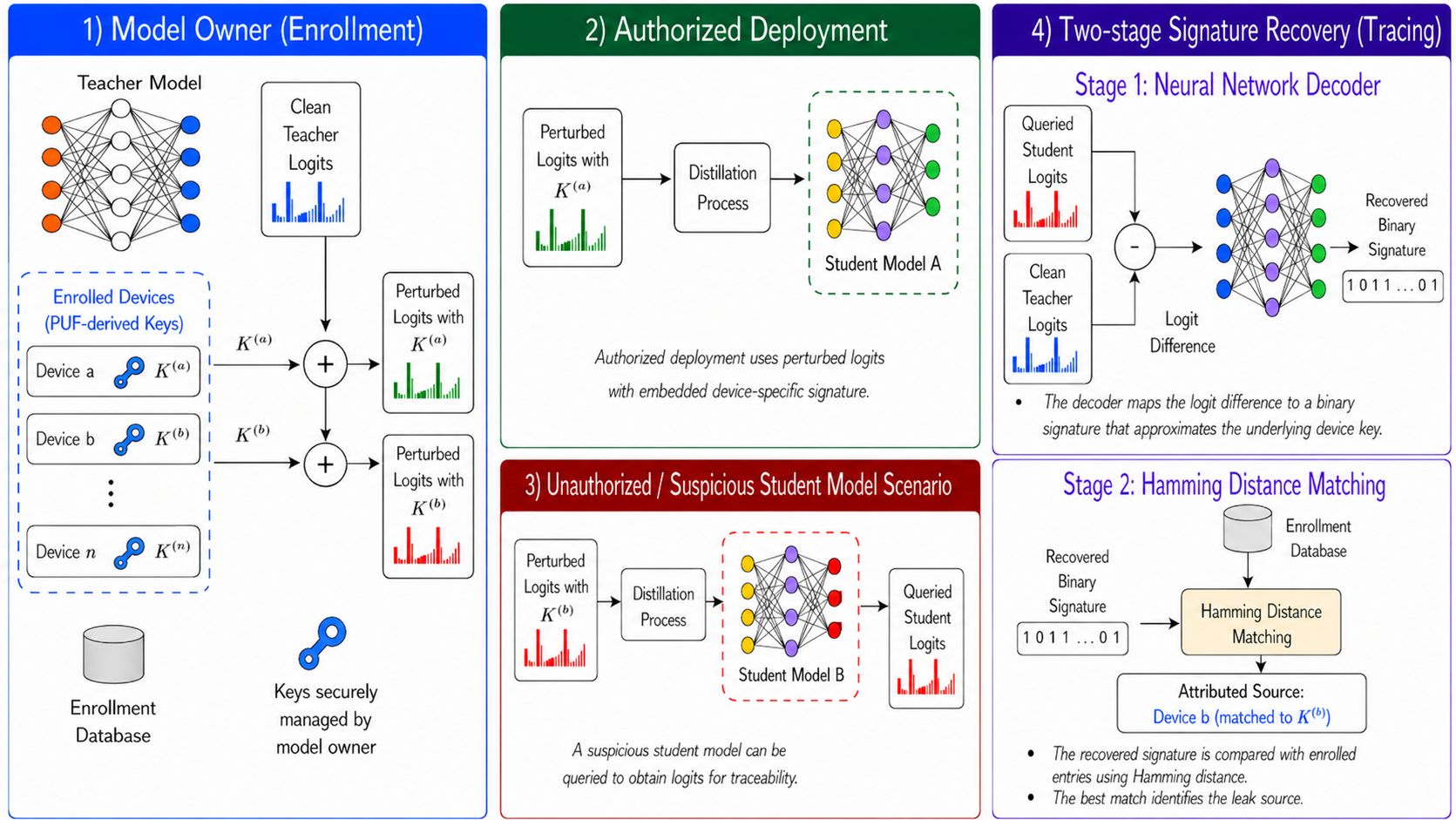}
\caption{System overview of the proposed hardware-rooted framework for device-level traceability in knowledge distillation.}
\label{fig:framework_overview}
\end{figure}

\section{Proposed Framework}
\label{sec:framework}

The proposed framework embeds a hardware-rooted identity into the behavioral outputs of deep learning models. As illustrated in Fig.~\ref{fig:framework_overview}, this enables post-leak traceability, allowing a stolen or distilled model to be reliably linked back to its originating device through output analysis.
\rev{Because the fingerprint is introduced at the model output level rather than inside architecture-specific internal layers, the same embedding mechanism can be applied to different model structures. Our experiments evaluate whether this fingerprint remains transferable and recoverable when the teacher and student architectures differ.}

\subsection{System Roles and Threat Model}
\label{subsec:roles}
To provide a rigorous foundation for accountability, we explicitly define the responsibilities and security boundaries of the three primary stakeholders in this ecosystem.


\textit{Model Owner (Defender)}:
The model owner serves as the central authority and defender. This entity is responsible for training the teacher model and managing its distribution to licensed hardware platforms. The owner maintains a \textbf{Secure Enrollment Database} containing the mapping between each authorized hardware instance and its enrolled reference PUF key $\mathbf{k}^{(m)}$. \rev{During traceback, the recovered signature is compared with these enrolled device-specific references using Hamming-distance matching, allowing limited PUF response variations to be tolerated.} The database is strictly internal and is used exclusively for ownership verification and device attribution. Importantly, traceback is an offline process performed solely by the owner and does not require the cooperation of the attacker.

\textit{Authorized PUF-Enabled Devices}:
Each authorized platform (e.g., an FPGA-based accelerator) contains an embedded PUF that generates a device-specific binary key $\mathbf{k}^{(m)} \in \{0,1\}^{n}$, where $m$ indexes the device and $n$ denotes the key length. The device uses its local PUF key to perturb the teacher model's output logits during operation. The key itself remains physically isolated within the hardware, ensuring that only its behavioral effect is visible in the output distribution.

\textit{Attacker (Distillation Adversary):} 
The adversary aims to steal the teacher model’s functionality through a distillation-based model theft attack. By gaining query access to the deployed teacher model, the attacker trains a surrogate student model to replicate its behavior. Crucially, the attacker aims to mimic the teacher's \textit{performance} but unknowingly transfers the \textit{PUF signature} into the student model's parameters. The attacker is assumed to have no access to the owner's enrollment database.

Instead of focusing on preventing theft, our framework is designed for post-leak traceability. We assume that an attacker may successfully obtain a functional student model; therefore, our primary objective is to ensure the model owner can still recover the embedded signature to identify the specific device that served as the source. To verify the robustness of this approach, we evaluate the system's performance against three adversarial strategies: In-Domain Fine-Tuning, Cross-Domain Transfer, and Logits/Latent Noise Injection. These tests confirm that the hardware identity remains recoverable even if an attacker attempts to clean the model by retraining it, repurposing it for a new task, or corrupting it with noise.

\subsection{FPGA-based RO-PUF Implementation}
\label{sec:ropuf_impl}

A practical evaluation must consider the non-ideal behavior of physical entropy sources. Therefore, we implement a hardware-rooted RO-PUF on a Xilinx Zynq-7020 (ZedBoard) FPGA to support robust end-to-end traceability. \rev{The proposed framework, however, is not FPGA-specific; it only requires a stable, device-specific PUF response. Therefore, the same embedding and recovery procedures can be extended to ASIC platforms using an appropriate on-chip PUF.}

In the proposed framework, \textbf{intra-device stability} is critical because the embedded fingerprint must remain reproducible under environmental variation for reliable recovery. \rev{The RO-PUF is selected for our FPGA prototype because its frequency-comparison mechanism provides a practical way to obtain a stable device-specific response.} Although latch-based PUFs have lower area overhead, their metastability-based operation makes them sensitive to transient noise. In contrast, RO-PUFs generate each bit through frequency comparison over a measurement window, where temporal integration smooths high-frequency jitter and improves reliability. The measurement window length (\texttt{MEAS\_CYC}) can be adjusted to further enhance stability without changing the RTL. \rev{Other PUF architectures, such as SRAM- or arbiter-based PUFs, could also be used provided that they offer sufficiently stable and distinctive device-specific responses.} Since mid-range FPGAs provide sufficient logic resources, we prioritize RO-PUF stability to maintain a consistent silicon-to-model identity mapping.


\begin{figure*}[t]
\centering
\includegraphics[width=\linewidth,height=5cm,keepaspectratio]{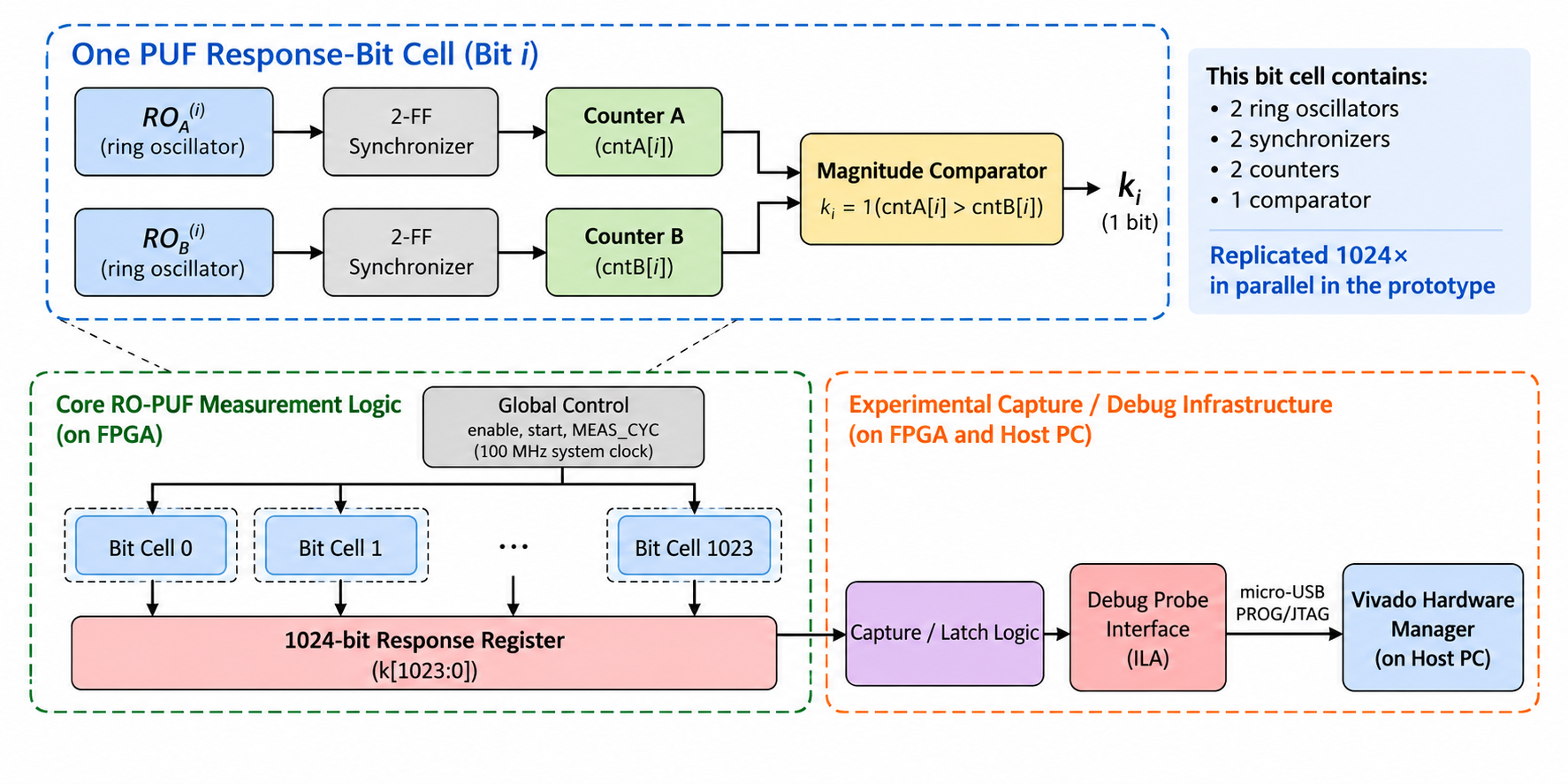}
\caption{\rev{Block diagram of the fully parallel 1024-bit RO-PUF experimental prototype. Each response-bit cell contains two ring oscillators, synchronization/counting logic, and a comparator, and the cell is replicated across the 1024-bit measurement array.}}
\label{fig:framework_ro}
\end{figure*}

\rev{The experimental platform implements a fully parallel 1024-bit RO-PUF pipeline ($n=1024$) operating at 100~MHz. Importantly, the 1024-bit response length is an experimental implementation choice rather than a requirement of the proposed fingerprinting framework. The required PUF key length depends on the output or latent dimension of the target model and the desired identity capacity; therefore, substantially shorter PUF responses may be sufficient in practical deployments. We use a 1024-bit implementation to obtain a large hardware-derived bit pool from which multiple non-overlapping shorter PUF keys can be selected for controlled device-identity experiments. Our goal is to construct multiple distinct logical PUF identities from the measured 1024-bit response so that the fingerprinting and attribution pipeline can be evaluated under different enrolled PUF keys. Because these shorter keys are derived from the measured RO-PUF response, they reflect hardware-dependent characteristics present in the prototype device, including bit bias, instability, and correlation. For example, a 1024-bit response can provide up to 51 non-overlapping 20-bit identities or 20 non-overlapping 50-bit identities. However, because all of these logical identities are derived from one physical FPGA, this construction does not constitute a full characterization of chip-to-chip inter-device variation.}

As illustrated in Figure~\ref{fig:framework_ro}, the prototype contains three main components: (i) a parallel RO-PUF measurement engine that generates response bits by comparing ring-oscillator frequencies over a predefined measurement interval, (ii) control and capture logic that coordinates measurement and stores the generated response, and (iii) an on-chip debug interface used for experimental observation and data collection. For hardware evaluation, the ZedBoard was connected to a host PC through the on-board PROG/JTAG interface, and Vivado Hardware Manager was used to capture the internal RO-PUF responses for offline analysis.

For each response bit $i \in [0,n-1]$, the measurement engine instantiates two independent ring oscillators, $RO_A^{(i)}$ and $RO_B^{(i)}$. Each oscillator is implemented as a reusable \texttt{ro\_osc} primitive, consisting of an odd-length inverter chain mapped to FPGA LUTs. The LUTs are annotated with \texttt{KEEP} and \texttt{DONT\_TOUCH} attributes to preserve the combinational feedback loop during synthesis, and a global \texttt{enable} signal controls oscillation.
When \texttt{enable} is asserted, the final inverter output is fed back to the first stage, forming a closed oscillating loop. When deasserted, the feedback node is forced to zero, breaking the loop and disabling oscillation.
Since the RO outputs are asynchronous, they pass through a two-stage register chain to reduce metastability. Saturating counters $\text{cntA}[i]$ and $\text{cntB}[i]$ count rising edges over the measurement window, and the response bit is generated as
$k_i = \mathbb{I}\big(\text{cntA}[i] > \text{cntB}[i]\big)$.
The resulting 1024-bit vector $\mathbf{k}$ is latched for downstream embedding. 

\rev{Because the prototype generates all 1024 response bits in parallel, the RO-pair, synchronization, counting, and comparison logic is replicated across the response-bit positions. Thus, the two ROs and two counters described for each response bit are instantiated repeatedly across the 1024-bit measurement array rather than shared by the entire PUF. The reported FPGA utilization also includes response storage, control/capture logic, and the on-chip debugging infrastructure used during hardware measurement. Therefore, the utilization reported in Table~\ref{tab:ro_puf_impl} represents the complete fully parallel 1024-bit experimental prototype rather than the minimum hardware overhead required by the proposed fingerprinting framework. A deployment using a shorter PUF response would require fewer replicated measurement resources; alternatively, resource sharing can trade additional response-generation latency for lower area. The goal of this implementation is to provide a stable and observable hardware PUF source for end-to-end validation rather than to demonstrate an area-optimal RO-PUF architecture.}


\begin{table}[t]
\centering
\caption{Post-Placement Resource Utilization of the 1024-bit RO-PUF Experimental Prototype on xc7z020}
\label{tab:ro_puf_impl}
\renewcommand{\arraystretch}{1.15}
\begin{tabular}{lccc}
\hline
\textbf{Resource} & \textbf{Used} & \textbf{Available} & \textbf{Utilization (\%)} \\
\hline
Slice LUTs & 36,210 & 53,200  & 68.1 \\
Registers  & 30,300 & 106,400 & 28.5 \\
BRAM Tiles & 15.5   & 140     & 11.1 \\
BUFGCTRL   & 3      & 32      & 9.4 \\
\hline
\end{tabular}
\begin{minipage}{0.95\columnwidth}
\footnotesize
\textit{Note:} Utilization is reported for the parallel 1024-bit experimental prototype, including response generation, control/capture, and on-chip debugging logic. 
\end{minipage}
\end{table}

\rev{Using measured RO-PUF responses allows the evaluation to incorporate real hardware-derived effects from the prototype FPGA, including RO delay variation, response instability, bit bias, and correlation. By partitioning the measured 1024-bit RO-PUF response into non-overlapping shorter key groups, we construct multiple distinct logical PUF identities to represent multiple enrolled device identities in the fingerprinting and recovery experiments. A broader evaluation across multiple FPGA devices remains important for characterizing population-level inter-device uniqueness and variation.}


\subsection{\rev{PUF-Based Fingerprinting Across Heterogeneous Architectures}}
\label{subsec:puf_fingerprinting}

\rev{The proposed fingerprint is embedded into the behavioral output of the teacher rather than into architecture-specific internal layers. This allows the same embedding mechanism to be applied to different model structures. The key question is therefore not whether cross-architecture distillation is possible, but whether the embedded PUF perturbation can preserve the teacher's predictive behavior while remaining sufficiently detectable to be inherited and recovered from the student. The perturbation magnitude is controlled by $\epsilon$, which determines the trade-off between model utility and fingerprint recoverability.}

During knowledge distillation, the student is trained to match the perturbed output distribution generated by the teacher and can therefore inherit the hardware identity associated with the originating device. As a result, a successfully distilled student model may carry a hardware-rooted fingerprint that can later be recovered for source attribution.

Each hardware device is represented by a binary PUF key $\mathbf{k} \in \{0,1\}^{n}$ generated by its local hardware PUF. To make this hardware identity compatible with neural network outputs, the binary key is converted into a bipolar signature vector $\mathbf{s} \in \{-1,+1\}^{n}$ using
\begin{equation}
s_i = 1 - 2k_i .
\label{eq:bipolar}
\end{equation}

Let $\mathbf{z}_t \in \mathbb{R}^{d}$ denote the clean output logits of the teacher model for an input sample, where $d$ represents the output dimension of the model (e.g., the number of classes in classification tasks). The hardware-conditioned teacher target is generated by superimposing the PUF signature onto the teacher output:
\begin{equation}
\mathbf{z}_{\text{PUF}} =
\mathbf{z}_t + \epsilon (\mathbf{U} \cdot \mathbf{s})
\label{eq:perturbed_logits}
\end{equation}
where $\epsilon$ controls the perturbation strength. The matrix $\mathbf{U} \in \mathbb{R}^{d\times n}$ is an optional orthonormal projection matrix that distributes the hardware entropy across the model output space when the PUF key length smaller than the output dimension. When $n = d$, the projection matrix is unnecessary and the perturbation can be applied directly as $\epsilon \mathbf{s}$. By keeping $\epsilon$ small relative to the natural variance of the logits, the primary prediction behavior of the model is preserved while embedding a structured hardware signature into the output distribution.

\textit{Categorical Fingerprinting (CNNs and Transformers)}:
For supervised classification models, the PUF signature is embedded at the logit layer. This design applies uniformly to both CNNs and transformer-based architectures, such as ViTs. This output-level embedding preserves the architecture-agnostic property of the proposed framework and enables consistent fingerprint propagation under model distillation or structural modification.

Since the fingerprint is introduced at the model output level, a student model can inherit the hardware identity even when its architecture differs from the teacher. For example, a CNN student distilled from a ViT teacher can still reproduce the embedded hardware signature. The student model is trained using a standard knowledge distillation objective:
\begin{equation}
\mathcal{L}_{\text{total}} =
(1-\alpha)\mathcal{L}_{\text{CE}}(y_s, y_{\text{true}})
+
\alpha\tau^2
\mathcal{L}_{\text{KD}}
\left(
\sigma\left(\frac{z_s}{\tau}\right),
\sigma\left(\frac{z_{\text{PUF}}}{\tau}\right)
\right)
\label{eq:classifier_loss}
\end{equation}
where $\mathcal{L}_{\text{CE}}(\cdot)$ denotes the cross-entropy loss between the student prediction $y_s$ and the ground-truth label $y_{\text{true}}$. The second term $\mathcal{L}_{\text{KD}}(\cdot)$ represents the knowledge distillation loss, which encourages the student to match the softened output distribution of the teacher. In this formulation, $z_s$ denotes the student logits and $z_{\text{PUF}}$ denotes the perturbed teacher logits containing the embedded hardware signature. The operator $\sigma(\cdot)$ represents the softmax function, while $\tau$ is the temperature parameter used to smooth the logit distributions during distillation. The weighting coefficient $\alpha \in [0,1]$ balances the contribution between the supervised learning objective and the distillation objective. Through this optimization process, the student model learns to replicate both the predictive behavior of the teacher and the PUF embedded perturbations in the output distribution.

\textit{Latent-Space Fingerprinting (Encoder-to-Encoder)}:
The same fingerprinting mechanism can be extended to representation-learning systems where the model output is a feature embedding rather than class probabilities. For an encoder model, the teacher produces a latent representation
$
\mathbf{z}_T(x) = \mathcal{E}_T(x) \in \mathbb{R}^{d},
$
where $\mathcal{E}_T(\cdot)$ denotes the teacher encoder and $d$ is the embedding dimension. This scenario commonly arises in metric-learning systems such as face recognition, where models output fixed-dimensional feature embeddings (e.g., 512-dimensional representations).

In such systems, the embedding dimension and representation space are typically standardized across architectures, allowing encoder-to-encoder distillation to directly align the student embedding with the teacher representation. Under this setting, the device-specific PUF signature can be embedded into the teacher latent representation prior to distillation.

Unlike classifier distillation aligns class logits, encoder-based distillation operates directly on latent representations. The teacher encoder produces the hardware-conditioned embedding $\mathbf{z}_{T}^{(m)}(x)$, while the student encoder $\mathcal{E}_S(\cdot)$ is trained to reproduce this representation using a latent alignment objective:
\begin{equation}
\mathcal{L} =
\lambda_{\mathrm{lat}}
\left\|
\mathcal{E}_S(x) - \mathbf{z}_{T}^{(m,w)}(x)
\right\|_2^2.
\label{eq:encoder_loss}
\end{equation}
This encoder-to-encoder distillation process transfers the embedding space from the teacher to the student model. Consequently, the embedded PUF signature becomes an intrinsic component of the learned representation, enabling device attribution by decoding the fingerprint from the recovered embedding.

To accurately reflect the constraints of FPGA-based edge accelerators, the teacher model is quantized to $q$ bits (e.g., $q=8$). We apply uniform quantization over weights, activations, and inputs to simulate the integer-only arithmetic pipelines used in standard deployments. For a real value $x \in [x_{\min}, x_{\max}]$, the quantized value $\hat{x}$ is calculated as:

\begin{equation}
    \hat{x} = \frac{1}{S} \cdot \text{round}\left(S \cdot \text{clip}(x, x_{\min}, x_{\max})\right), 
    \label{eq:quantization}
\end{equation}
where $S = \frac{2^q - 1}{x_{\max} - x_{\min}}$ is the scaling factor. The teacher model is fine-tuned under these quantization-aware constraints to maintain its baseline utility. The signature perturbation $\epsilon \cdot (\mathbf{U} \cdot \mathbf{s})$ is applied to these quantized logits, ensuring the framework's compatibility with low-power hardware accelerators.

\subsection{Scalable Multi-Level Logit Encoding Scheme}
\label{sec:bit_compression}

A primary constraint in behavioral model fingerprinting arises from the limited dimensionality of the model output space. In the simplest formulation, each PUF bit is mapped to one logit dimension ($n=d$), resulting in a one-to-one correspondence between the PUF key and the output vector. Under this mapping, the theoretical identifier space is bounded by $2^d$. In practice, the effective identifier capacity is even smaller due to entropy loss, bit-bias, and environmental instability in real PUF measurements, as discussed in Section~\ref{sec:puf_background}. Consequently, such a direct mapping becomes insufficient for large-scale industrial deployments where millions or billions of unique device identifiers may be required.

To address this limitation, we introduce a \textbf{Multi-Level Logit Encoding} strategy that increases the information density carried by the model output. Instead of assigning a single PUF bit to each logit dimension, the proposed scheme treats the teacher's logits as a multi-level communication channel, allowing each logit to encode $e$ bits of PUF entropy. This design effectively increases the fingerprint capacity without modifying the model architecture. As a result, the total identity space expands exponentially to $2^{e \cdot d}$, enabling scalable device-level fingerprinting for large deployment scenarios.
In this formulation, the $n$-bit PUF key $\mathbf{k}$ is partitioned into $d$ segments, where each segment contains $e$ bits. Each segment is mapped to a single logit dimension $z_i$. To implement this, we interpret each $e$-bit segment $(b_0, b_1, \dots, b_{e-1})$ as a signed integer $I_i$ using standard two's complement representation:
\begin{equation}
I_i \in \{-2^{e-1}, \dots, 2^{e-1} - 1\}.
\end{equation}
The integer value $I_i$ is then linearly transformed into a logit perturbation $\delta_i$ via 
$\delta_i = \epsilon \cdot (I_i + 0.5),
\label{eq:compression_mapping}
$
where $\epsilon$ is a hyperparameter representing the perturbation strength. The $+0.5$ shift ensures the hardware signature is balanced around zero, preventing it from constantly pushing the model's outputs in one direction.
Table~\ref{tab:bitmap-example} illustrates an example using a 3-bit-per-logit configuration ($e=3$). This mapping provides 8 discrete perturbation levels for each logit dimension with a perturbation strength of $\epsilon=0.4$.
\begin{table}[h]
\centering
\caption{Encoding Mapping for $e=3$ ($\epsilon = 0.4$)}
\label{tab:bitmap-example}
\begin{tabular}{ccc}
\hline
Binary Segment ($\mathbf{b}_i$) & Integer ($I_i$) & Perturbation ($\delta_i$) \\
\hline
000 & 0 & $+0.2$ \\
001 & 1 & $+0.6$ \\
010 & 2 & $+1.0$ \\
011 & 3 & $+1.4$ \\
100 & -4 & $-1.4$ \\
101 & -3 & $-1.0$ \\
110 & -2 & $-0.6$ \\
111 & -1 & $-0.2$ \\
\hline
\end{tabular}
\end{table}

\rev{The computational overhead of the multi-level encoding is confined to the fingerprinted teacher side. For a $d$-dimensional output and an $n=e\cdot d$-bit PUF key, each $e$-bit segment is mapped to one perturbation level and added to the corresponding teacher logit. Thus, the encoding requires only $d$ bit-to-level mappings and $d$ element-wise additions after the teacher forward pass, without introducing additional neural-network layers or extra forward passes. These operations are independent across output dimensions and can be performed in parallel. The more implementation-dependent overhead arises from generating the $n=e\cdot d$ PUF bits. In a parallel hardware implementation, PUF response generation can potentially overlap with the teacher inference pipeline, while a more serialized implementation can reduce hardware cost at the expense of additional latency.}

The primary advantage of this multi-level encoding is its exponential scalability relative to the encoding density $e$.
While increasing $e$ significantly expands the identity space, it also reduces the spacing between adjacent encoding levels in the logit space. Assuming the perturbation values are uniformly distributed within the interval $[-\epsilon,\epsilon]$, the minimum spacing between two neighboring levels can be expressed as
$
\Delta = \frac{2\epsilon}{2^e - 1}.
$
As $e$ increases, the number of encoding levels grows exponentially, while the spacing $\Delta$ correspondingly decreases. Consequently, even small deviations in the student outputs caused by distillation noise may shift a perturbed logit across neighboring decision boundaries, increasing the probability of decoding errors.

\rev{This introduces a practical capacity--utility--robustness trade-off. Increasing $e$ allows more PUF bits to be encoded per logit, but the denser encoding requires sufficient perturbation strength to remain distinguishable after distillation. The operating point is therefore selected to provide reliable fingerprint recovery while keeping the student-model utility close to its unencoded baseline.}

\rev{The two-stage recovery pipeline described in Section~\ref{sec:recovery} further improves robustness to decoding errors. The neural decoder estimates the embedded PUF key from the distorted student outputs, while the Stage-II Hamming-distance matcher can tolerate a limited number of recovered-bit errors by selecting the nearest enrolled identity.}
By allowing multiple PUF bits to be encoded per output dimension, the scheme increases identity capacity without requiring a proportional increase in the teacher output dimension.



\begin{algorithm}[t]
\caption{Synthetic Logit Simulation and Neural Decoder Training}
\label{alg:decoder_train}
\begin{algorithmic}[1]
\State \textbf{Input:} Master PUF bitstream $\mathbf{B}_{raw}$, key length $n$, logit dimension $d$, perturbation set $\mathcal{E}$, noise level $\sigma$, device registry size $M$, optional projection matrix $\mathbf{U}$
\State \textbf{Output:} Trained MLP Identity Decoder $\mathcal{D}_\theta$

\State \textbf{Step 1: Synthetic Data Initialization}
\State Extract $M$ unique keys from $\{0,1\}^n$ to form subset $\mathcal{K} = \{\mathbf{k}_1, \dots, \mathbf{k}_M\}$
\State Initialize training dataset $\mathcal{S} = \emptyset$

\State \textbf{Step 2: Synthetic Logit Simulation}
\For{each device key $\mathbf{k}_i \in \mathcal{K}$}
    \State Randomly sample perturbation scale $\epsilon \sim \mathcal{E}$
    \State Map bits to bipolar signs: $\mathbf{s}_i = (1 - 2\mathbf{k}_i)$
    
    \State \textbf{if} $\mathbf{U}$ is provided: $\boldsymbol{\delta}_i = \epsilon \cdot (\mathbf{U} \cdot \mathbf{s}_i)$ \textbf{else:} $\boldsymbol{\delta}_i = \epsilon \cdot \mathbf{s}_i$
    
    \For{sample $j = 1$ to $Q$}
        \State Generate clean teacher logits: $\mathbf{z}_{t}^{(j)} \sim \mathcal{N}(0, \mathbf{I}_d)$
        \State Perturb teacher logits: $\mathbf{z}_{\text{PUF}}^{(j)} = \mathbf{z}_{t}^{(j)} + \boldsymbol{\delta}_i$
        \State Simulate student behavior: $\mathbf{z}_{s}^{(j)} = \mathbf{z}_{\text{PUF}}^{(j)} + \boldsymbol{\eta}\quad $
        \State Compute logit residual: $\Delta \mathbf{z}^{(j)} = \mathbf{z}_{s}^{(j)} - \mathbf{z}_{t}^{(j)}$
        \State $\mathcal{S} \gets \mathcal{S} \cup \{(\Delta \mathbf{z}^{(j)}, \mathbf{k}_i)\}$
    \EndFor
\EndFor

\State \textbf{Step 3: Neural Decoder Training and Optimazation}
\State Train MLP decoder $\mathcal{D}_\theta$ on $\mathcal{S}$ using BCE loss
\State Use sigmoid activation and early stopping to optimize parameters $\theta$
\State \Return Trained Identity Decoder $\mathcal{D}_{\theta}$
\end{algorithmic}
\end{algorithm}

\subsection{Two-Stage Signature Recovery}
\label{sec:recovery}

\rev{Once a suspicious distilled student model becomes available for analysis or black-box querying, the model owner can recover its inherited PUF fingerprint using the proposed two-stage recovery process. The defender does not need physical access to the adversary’s device. The PUF-enabled hardware is only used when the teacher generates device-specific outputs during distillation. During verification, the suspicious student model is queried, the embedded fingerprint is recovered, and the result is compared with the enrolled PUF keys to identify the associated hardware instance. If the suspicious student model is completely inaccessible, post-leak attribution cannot be performed.} As shown in Fig.~\ref{fig:framework_overview}, we propose a two-stage recovery process that combines the pattern-recognition capabilities of deep learning with the rigorous error-correction of nearest-neighbor decoding. 

\textit{Stage I: Neural Network-Based Recovery}. The first stage of the verification process involves a neural decoder $\mathcal{D}_\theta$ designed to map behavioral model deviations back to their underlying hardware-rooted keys. A primary challenge in training a neural decoder is the unavailability of real student models for every possible PUF key in the space $\mathcal{K}$. To overcome this, we utilize a Synthetic Logit Simulation approach to generate a robust training dataset without requiring expensive distillation cycles. Because the decoder does not require the specific internal weights of the student model; instead, it learns to distinguish the structured perturbation $\boldsymbol{\delta}$ from the combined noise of the teacher's quantization and the student's learning variance.

As detailed in Algorithm~\ref{alg:decoder_train}, we construct a training set $\mathcal{S}$ by simulating the statistical relationship between a teacher model's output and a distilled student's behavior. The simulation follows three main steps:

\begin{enumerate}
    \item \textbf{Synthetic Data Initialization:} Extract $M$ unique $n$-bit hardware keys, denoted as $\mathcal{K} = \{\mathbf{k}_1, \dots, \mathbf{k}_M\}$.
    \item \textbf{Synthetic Logit Simulation:} For each device, we model the student's output $\mathbf{z}_{s}$ as a stochastic function of the teacher's clean logits $\mathbf{z}_{t}$. We generate synthetic teacher logits following a standard normal distribution, $\mathbf{z}_{t} \sim \mathcal{N}(0, \mathbf{I}_d)$.
    We map the binary key to bipolar signs $\mathbf{s}_i$ and compute a perturbation vector $\boldsymbol{\delta}_i$. If the projection matrix $\mathbf{U}$ is provided (to handle cases where $n \neq d$), the signature is defined as $\boldsymbol{\delta}_i = \epsilon \cdot (\mathbf{U} \mathbf{s}_i)$; otherwise, it is directly applied as $\boldsymbol{\delta}_i = \epsilon \cdot \mathbf{s}_i$. 
    Therefore, the simulated student behavior is then expressed as:
    \begin{equation}
\mathbf{z}_{\text{s}}^{(j)} = \mathbf{z}_t^{(j)} + \boldsymbol{\delta}_i + \boldsymbol{\eta}, \quad \boldsymbol{\eta} \sim \mathcal{N}(\mathbf{0}, \sigma^2 \mathbf{I}_d)
\label{eq:simulation_model}
    \end{equation}
    where $\boldsymbol{\eta}$ represents the approximation noise inherent in the distillation process. By calculating the logit residual $\Delta \mathbf{z} = \mathbf{z}_{s} - \mathbf{z}_{t}$, the decoder is trained to isolate the hardware signature from the model’s original prediction information.

    \item \textbf{Neural Decoder Training and Optimization:}
    We implement $\mathcal{D}_\theta$ as a Multi-Layer Perceptron (MLP) optimized for multi-label classification. 
\end{enumerate}
Given a residual vector $\Delta \mathbf{z}$, the decoder outputs a probability vector $\mathbf{p} \in [0, 1]^n$.
    The objective is to minimize the Bitwise Binary Cross-Entropy (BCE) loss:
    \begin{equation}
    \mathcal{L}_{\text{BCE}} = - \frac{1}{n} \sum_{b=1}^{n} \left[ k_{i,b} \log p_{b}^{(j)} + (1 - k_{i,b}) \log (1 - p_{b}^{(j)}) \right]
    \end{equation}
    where $k_{i,b}$ denotes the ground-truth value of the $b$-th bit of the PUF key. Each output dimension $p_b^{(j)}$ represents the predicted probability that the $b$-th PUF bit equals 1. 
    After inference, the recovered PUF bits are obtained through a threshold operation on the predicted probabilities:
    \begin{equation}
    \hat{k}_{b} =
    \begin{cases}
    1, & p_b \ge 0.5 \\
    0, & \text{otherwise}.
    \end{cases}
    \end{equation}
    The resulting binary vector $\hat{\mathbf{k}} \in \{0,1\}^n$ represents the recovered PUF key.

\textit{Stage II: Hamming Distance decoder.}
The output of the neural decoder is a predicted binary string $\hat{\mathbf{k}}$. However, due to the $1\%$--$5\%$ intra-device bit flips characteristic of physical hardware (as discussed in Section~\ref{sec:puf_background}) and the stochastic nature of the distillation process, $\hat{\mathbf{k}}$ may contain localized bit errors. To provide deterministic accountability, we apply a second-stage Hamming distance decoder. This stage acts as a nearest neighbor estimator by comparing the predicted string $\hat{\mathbf{k}}$ against the model owner's internal enrollment registry database $\mathcal{K}$. The final recovered identity $\mathbf{k}^*$ is defined as a registered key that minimizes the Hamming distance:
$\mathbf{k}^* = \arg\min_{\mathbf{k} \in \mathcal{K}} d_H(\hat{\mathbf{k}}, \mathbf{k}),$
where $d_H(\cdot, \cdot)$ denotes the Hamming distance.

 This two-stage approach enhances robustness to bit errors by combining neural decoding with a nearest-neighbor search in the authorized database, enabling reliable attribution even when the recovered signature is imperfect.  The neural decoder first extracts the hidden signal from the high-dimensional logits, while the Hamming distance decoder then refines the result by correcting small errors. This two-stage design makes the recovery process robust to noise from the student model and stable even under variations in hardware-based PUF responses.

\section{Experimental Evaluation}
\label{sec:experiments}

This section presents a comprehensive quantitative evaluation of the proposed PUF-based fingerprinting framework. We systematically assess its ability to trace distilled models back to their originating hardware device under varying classification complexities, architectural paradigms, and identity-space scales, demonstrating its effectiveness for  device-level traceability.

\subsection{Experimental Setup and Methodology}
\label{subsec:setup}

The unique fingerprints for our system are taken from a RO-PUF built on a Xilinx Zynq-7020 (ZedBoard) FPGA. We collected a 1024-bit RO-PUF response under standard operating conditions ($25^\circ\text{C}$, 1.0V). This response is used as the physical entropy source for the identity registry. For large-scale evaluation, we partition the measured 1024-bit RO-PUF response into non-overlapping \(k\)-bit segments to construct multiple hardware-derived candidate identities for evaluating the multi-identity recovery pipeline
By using this hardware-aware simulation, we ensure that the identity embedded in the model remains easy to verify even when hardware noise occur.

To prove the framework's \textbf{architecture-agnostic} versatility, we evaluate three distinct ``Teacher-Student'' couplings. These configurations, detailed in Table~\ref{tab:arch_config}, represent significant structural shifts, testing the fingerprint's survival across global attention mechanisms, local convolutions, and latent-space projections.

\begin{itemize}
    \subsubsection{Encoder-to-Encoder Workflow (CIFAR-10)}
\label{subsubsec:enc_to_enc}

    \item \textbf{Enc-to-Enc (CIFAR-10):} we evaluate the framework's efficacy within the paradigm of representation learning. While the teacher is a full Symmetric Autoencoder (AE) capable of image reconstruction, the student is implemented as a resource-constrained model by using case 2 in Section~\ref{subsec:puf_fingerprinting} with latent dimension 64 and PUF bit length 50. The encoder backbone utilizes a sequence of strided and standard convolutions: $32c3(s2) \to 64c3(s2) \to 64c3 \to F64$, mapping the input to a 64-dimensional latent bottleneck. To establish the baseline manifold, the teacher employs a mirrored decoder: $F4096 \to 64c3 \to U2 \to 32c3 \to U2 \to 3c3$.
    To ensure that the hardware-rooted signature survives the distillation process,  the loss incorporates a critical latent alignment term ($\text{MSE}(z_s, z_{t,q,w})$) as shown in Eq.~\ref{eq:encoder_loss}. This objective forces the student's latent space to replicate the teacher’s fingerprinted and quantized manifold.

    \item \textbf{Trans-to-CNN (CIFAR-20):} We investigate heterogeneous distillation by transferring knowledge from a TinyViT-21M model to a customized SmallCNN on CIFAR-20. The TinyViT-21M teacher uses a 192-dimensional feature representation and 6 multi-head self-attention (MHSA) heads. The student is designed as a compact convolutional network with the following architecture: $64c3 \to 128c3(s2) \to 256c3(s2) \to 256c3 \to GAP \to F192 \to F20$, where  $GAP$ denotes Global Average Pooling. $F192$ is a 192-dimensional embedding layer, and $F20$
   is the final classifier for the 20 classes.
    
    \rev{\item \textbf{CNN-to-Transformer (CIFAR-20):}
To evaluate the reverse direction of heterogeneous distillation, we additionally
distill a CNN teacher into a compact Vision Transformer student on CIFAR-20.
The CNN teacher contains approximately 3.46M parameters, while the Transformer
student contains approximately 0.387M parameters. The student uses a patch size
of 4, a 96-dimensional embedding, four Transformer blocks, and four
self-attention heads, followed by a 20-class classification head. This
configuration complements the Transformer-to-CNN setting and evaluates whether
the hardware-rooted fingerprint can remain recoverable when the architectural
ordering of the teacher and student is reversed.}

    \item \textbf{CNN-to-CNN (CIFAR-50):} 
   The teacher model is a compact VGG-style CNN (Mini-VGG), derived from the stacked
 convolutional design of VGG~\cite{simonyan2015vgg}. Specifically, the teacher follows a structural sequence of $(64c3 \times 2) \to \text{MP} \to (128c3 \times 2) \to \text{MP} \to (256c3 \times 2) \to \text{GAP} \to F256 \to F50$, where $MP$ denotes MaxPooling.  This high-capacity teacher is distilled into a highly compressed student backbone: $64c3(s2) \to 128c3(s2) \to 128c3(s2) \to F256 \to F50$. This configuration is specifically designed to test whether the hardware-rooted signature can survive significant spatial information loss and structural simplification during the transition from a deep, block-based network to a shallow network.
    
\end{itemize}

\begin{table*}[t]
\centering
\small
\caption{System Configurations and Detailed Architectural Mappings for Identity Verification}
\label{tab:arch_config}

\begin{tabular*}{\textwidth}{@{\extracolsep{\fill}}lllc@{}}
\toprule
\textbf{Workflow} &
\textbf{Teacher Structure} &
\textbf{Student Structure} &
\textbf{Dataset} \\
\midrule

\textbf{Enc-to-Enc}
& Symmetric AE
& $3c3$-$2s$-$32$-$32c3$-$2s$-$64$-$64c3$-$F64$
& CIFAR-10 \\

\midrule

\textbf{Trans-to-CNN}
& TinyViT: $MHSA(6h)$+$MLP$
& $64c3$-$128c3(s2)$-$256c3(s2)$-$256c3$-$F192$-$F20$
& CIFAR-20 \\

\midrule

\textbf{CNN-to-Trans}
& CNN ($\sim$3.46M params)
& ViT: $P4$-$F96$-$[MHSA(4h)$+$MLP]_{\times4}$-$F20$
& CIFAR-20 \\

\midrule

\textbf{CNN-to-CNN}
& 8-layer CNN
& $64c3(s2)$-$128c3(s2)$-$128c3(s2)$-$F256$-$F50$
& CIFAR-50 \\

\bottomrule
\end{tabular*}
\end{table*}


We define three core metrics to quantify the trade-off between model utility and recovery reliability:
\begin{itemize}
    \item \textbf{Baseline and Fingerprinted Utility:} These metrics evaluate the functional performance of the student model in two states: the \textit{Baseline} (the model after standard distillation) and the \textit{Fingerprinted} (the model after the PUF identity has been embedded). The specific metric used depends on the workflow: For the \textbf{Enc-to-Enc} workflow, utility is quantified via Mean Squared Error (MSE). In this context, lower values represent higher reconstruction fidelity. For the \textbf{Trans-to-CNN} and \textbf{CNN-to-CNN} workflows, utility is quantified via Top-1 Classification Accuracy (\%), where higher values signify superior task performance.

    \item \textbf{Bit Error Rate (BER):} Measures the fraction of incorrect bits in reconstructed keys. $BER = \frac{1}{n} \sum_{i=1}^{n} |b_i - \hat{b}_i|$
    where $n$ is the key length, $b$ is the enrolled bit, and $\hat{b}$ is the bit reconstructed.
    \item \textbf{Frame Error Rate (FER):} Represents the probability that the recovered key is not correct. 
    A final FER of 0\% indicates that the identity was traced with absolute certainty back to the specific source device.
\end{itemize}

\rev{ We evaluate representative values of the fingerprinting-strength parameter
$\epsilon$ to examine the trade-off between student utility and fingerprint
recoverability. The selected values are intended to capture an unreliable
recovery regime, the smallest evaluated value that achieves reliable recovery,
and a larger value used to examine whether further increasing the fingerprint
strength provides additional recovery benefit.}

\vspace{-10pt}
\subsection{Impact of Fingerprint Strength on Model Performance and Traceability}
\label{subsec:strength_impact}


The effectiveness of fingerprinting depends on the perturbation magnitude
$\epsilon$, which controls the \textbf{traceability--utility trade-off}.
To quantify this trade-off, we measure the performance of a clean baseline
student and the corresponding fingerprinted student model. For the
classification tasks (CIFAR-20 and CIFAR-50), utility is measured by
top-1 accuracy, while for the encoder task (CIFAR-10), utility is measured
by MSE. Traceability is evaluated using the final BER and FER after
Stage-II Hamming-distance refinement.

\rev{Table~\ref{tab:final_performance} reports representative operating points
rather than a fine-grained characterization of the complete
BER/FER-versus-$\epsilon$ curve. For each workflow, we identify the
smallest \emph{evaluated} $\epsilon$ at which both final BER and FER reach
zero while maintaining acceptable student-model utility, and we additionally
report a larger $\epsilon$ value to examine whether further increasing the
fingerprinting strength provides additional recovery benefit. The reported
values should therefore be interpreted as practical evaluated operating
points rather than exact theoretical thresholds.}
\begin{itemize}
    \item \textbf{Enc-to-Enc (CIFAR-10):}
    At $\epsilon=0.005$, the identity remains poorly resolved, with a final BER of $12.21\%$ and a final FER of $25.0\%$. Among the evaluated values, $\epsilon=0.010$ is the smallest reliable operating point, achieving $0.0\%$ final BER and FER. Increasing $\epsilon$ further to $0.020$ maintains zero-error recovery but provides no additional recovery benefit.

    \item \textbf{Trans-to-CNN (CIFAR-20):}
    At $\epsilon=0.05$, recovery remains unreliable, with a final BER of $15.62\%$ and a final FER of $43.75\%$. Among the evaluated values, $\epsilon=0.10$ is the smallest value that achieves $0\%$ final BER and FER. Increasing $\epsilon$ to $0.25$ maintains the same zero-error recovery performance.
    \rev{\item \textbf{CNN-to-Transformer (CIFAR-20):}
At $\epsilon=0.05$, recovery remains unreliable, with a final BER of
$19.17\%$ and a final FER of $50.0\%$. Among the evaluated perturbation
strengths, $\epsilon=0.09$ is the smallest value that achieves $0\%$ final BER
and $0\%$ final FER, corresponding to $100\%$ device recovery accuracy.
Increasing $\epsilon$ to $0.25$ maintains the same zero-error final recovery.
The student classification accuracy remains approximately $50\%$ across these
evaluated operating points.}

    \item \textbf{CNN-to-CNN (CIFAR-50):}
    At $\epsilon=0.020$, recovery remains unreliable, with a final BER of $23.0\%$ and a final FER of $45.0\%$. Among the evaluated values, $\epsilon=0.05$ is the smallest value that achieves $0\%$ final BER and FER, while $\epsilon=0.10$ maintains zero-error recovery. The measured student accuracy at $\epsilon=0.05$ is slightly higher than the baseline in this experiment, but this utility increase is treated as an empirical observation rather than a required effect of the fingerprinting mechanism.
\end{itemize}

\begin{table*}[t]
\caption{Model Utility and Final Identity Traceability at Representative Fingerprint Strengths}
\label{tab:final_performance}
\centering
\small
\setlength{\tabcolsep}{3pt}
\renewcommand{\arraystretch}{1.12}

\begin{tabularx}{\textwidth}{@{}L L c c C c c c@{}}
\toprule
\multirow{2}{*}{\textbf{Workflow}}
& \multicolumn{2}{c}{\textbf{Baseline}}
& \multirow{2}{*}{\textbf{$\epsilon$}}
& \multirow{2}{*}{\shortstack[c]{\textbf{Fingerprinted}\\\textbf{Value}}}
& \multirow{2}{*}{\shortstack[c]{\textbf{Final}\\\textbf{BER (\%)}}}
& \multirow{2}{*}{\shortstack[c]{\textbf{Final}\\\textbf{FER (\%)}}}
& \multirow{2}{*}{\shortstack[c]{\textbf{Trace-}\\\textbf{ability}}} \\
\cmidrule(lr){2-3}
& \textbf{Metric} & \textbf{Value} & & & & & \\
\midrule

\multirow{3}{*}{%
\shortstack[l]{\textbf{Enc-to-Enc}\\(CIFAR-10)}}
& MSE ($\times 10^{-3}$) & 7.45
& 0.005 & $7.49 \pm 0.028$ & 12.21 & 25.0 & Low \\

& & &
0.01 & \textbf{$7.11 \pm 0.084$}
& \textbf{0.00} & \textbf{0.0} & \textbf{High} \\

& & &
0.02 & $7.43 \pm 0.095$
& 0.00 & 0.0 & High \\
\midrule

\multirow{3}{*}{%
\shortstack[l]{\textbf{Trans-to-CNN}\\(CIFAR-20)}}
& Acc. (\%) & 70.81
& 0.05 & $70.76 \pm 0.33$ & 15.62 & 43.75 & Low \\

& & &
0.10 & \textbf{$70.83 \pm 0.34$}
& \textbf{0.00} & \textbf{0.0} & \textbf{High} \\

& & &
0.25 & $70.78 \pm 0.58$
& 0.00 & 0.0 & High \\
\midrule

\multirow{3}{*}{%
\shortstack[l]{\textbf{CNN-to-Trans}\\(CIFAR-20)}}
& Acc. (\%) & 50.74
& 0.05 & $50.21 \pm 0.37$ & 19.17 & 50.0 & Low \\

& & &
0.09 & \textbf{$50.18 \pm 0.66$}
& \textbf{0.00} & \textbf{0.0} & \textbf{High} \\

& & &
0.25 & $50.02 \pm 0.40$
& 0.00 & 0.0 & High \\
\midrule

\multirow{3}{*}{%
\shortstack[l]{\textbf{CNN-to-CNN}\\(CIFAR-50)}}
& Acc. (\%) & 52.40
& 0.02 & $52.40 \pm 1.34$ & 23.00 & 45.0 & Low \\

& & &
0.05 & \textbf{$53.17 \pm 1.56$}
& \textbf{0.00} & \textbf{0.0} & \textbf{High} \\

& & &
0.10 & $53.16 \pm 1.81$
& 0.00 & 0.0 & High \\
\bottomrule
\end{tabularx}

\vspace{1mm}
\footnotesize
\raggedright
\textit{Note:} For Enc-to-Enc, the baseline and fingerprinted values are MSE, where lower is better. For Trans-to-CNN, CNN-to-Transformer, and CNN-to-CNN, they are top-1 accuracy (\%), where higher is better. Final BER and FER are reported after the two-stage recovery process.
\end{table*}

The experimental results show that the student model can inherit the hardware signature when the fingerprinting strength is sufficiently large. For the Enc-to-Enc workflow, $\epsilon=0.005$ results in a final BER of $12.21\%$ and a final FER of $25.0\%$, whereas $\epsilon=0.010$ is the smallest evaluated value that achieves $0\%$ final BER and FER. Increasing the strength further to $\epsilon=0.020$ maintains zero-error recovery but provides no additional recovery benefit.

Similar behavior is observed for the Trans-to-CNN and CNN-to-CNN workflows. The reported operating point should therefore be interpreted as the smallest evaluated perturbation strength that achieves reliable recovery while maintaining acceptable model utility, rather than as an exact theoretical threshold. In practice, $\epsilon$ can be calibrated by evaluating candidate values and selecting the smallest value that satisfies the desired recovery reliability and utility requirements.




\vspace{-14pt}
\subsection{Two-Stage Recovery Efficiency}
\label{subsec:multi_stage_recovery}

While the perturbation magnitude $\epsilon$ dictates the primary signal quality, the ultimate reliability of the proposed framework depends on the transition from soft-bit estimation to hard-identity attribution. We evaluate this through a hierarchical two-stage recovery pipeline: (i) the \textbf{Neural Decoder Stage}, which isolates the signature from the logit residuals, and (ii) the \textbf{Hamming-Distance Refinement Stage}, which serves as the final identity attribution layer by identifying the closest matching device within the enrolled PUF registry.

\subsubsection{Stage-I: Neural Decoder Performance}

\begin{figure}[htbp]
\centering
\makebox[\columnwidth][c]{%
\begin{minipage}{0.6\columnwidth}
\centering
\begin{tikzpicture}[trim axis left]
\begin{axis}[
    ybar,
    bar width=5pt,
    width=\linewidth,
    height=0.6\linewidth,
    scale only axis,
    xmin=0, xmax=3.5,
    ymin=0, ymax=50,
    ylabel={BER(\%)},
    xtick={0.5, 2, 3.5},
    xticklabels={Enc-to-Enc, Trans-to-CNN, CNN-to-CNN},
    xticklabel style={font=\small, yshift=-5pt},
    ylabel style={font=\scriptsize},
    tick label style={font=\scriptsize},
    nodes near coords,
    every node near coord/.append style={font=\scriptsize},
    legend style={
        font=\scriptsize,
        at={(0.5,-0.25)},
        anchor=north,
        legend columns=2
    },
    axis lines=box,
    clip=false,
    enlarge x limits=0.2,
    ymajorgrids=true,
    xmajorgrids=true,
    grid style=dashed,
    ytick={10,20,30,40,50},
    extra x tick style={grid style=dashed},
]

\pgfmathsetmacro{\shift}{0.06}

\addplot+[draw=olive!80!black, fill=olive!30]
coordinates {(1 - 0.3*\shift, 23.0)};

\addplot+[draw=olive!80!black, fill=olive!30]
coordinates {(1 + 0.3*\shift, 4.2)};

\addplot+[draw=olive!80!black, fill=olive!30]
coordinates {(1 + 0.5*\shift, 0.0)};

\node[font=\tiny, anchor=north east, rotate=45]
at (axis cs:1 - 12*\shift, 1) {0.005};

\node[font=\tiny, anchor=north east, rotate=45]
at (axis cs:1 - 9*\shift, 1) {0.01};

\node[font=\tiny, anchor=north east, rotate=45]
at (axis cs:1 - 6*\shift, 1) {0.02};

\addplot+[draw=orange!80!black, fill=orange!40]
coordinates {(2 - 0.3*\shift, 21.9)};

\addplot+[draw=orange!80!black, fill=orange!40]
coordinates {(2 + 0.3*\shift, 8.7)};

\addplot+[draw=orange!80!black, fill=orange!40]
coordinates {(2 + 0.5*\shift, 0.0)};

\node[font=\tiny, anchor=north east, rotate=45]
at (axis cs:2 - 5*\shift, 1) {0.05};

\node[font=\tiny, anchor=north east, rotate=45]
at (axis cs:2 - 1*\shift, 1) {0.10};

\node[font=\tiny, anchor=north east, rotate=45]
at (axis cs:2 + 2*\shift, 1) {0.25};

\addplot+[draw=teal!80!black, fill=teal!30]
coordinates {(3 - 0.5*\shift, 40.3)};

\addplot+[draw=teal!80!black, fill=teal!30]
coordinates {(3 - 0.3*\shift, 29.5)};

\addplot+[draw=teal!80!black, fill=teal!30]
coordinates {(3 + 0.3*\shift, 8.44)};

\addplot+[draw=teal!80!black, fill=teal!30]
coordinates {(3 + 0.5*\shift, 0.8)};

\node[font=\tiny, anchor=north east, rotate=45]
at (axis cs:3 + 3*\shift, 1) {0.01};

\node[font=\tiny, anchor=north east, rotate=45]
at (axis cs:3 + 6*\shift, 1) {0.02};

\node[font=\tiny, anchor=north east, rotate=45]
at (axis cs:3 + 10*\shift, 1) {0.05};

\node[font=\tiny, anchor=north east, rotate=45]
at (axis cs:3 + 13*\shift, 1) {0.1};

\end{axis}
\end{tikzpicture}
\end{minipage}%
}

\caption{BER after first-stage recovery across different PUF lengths and perturbation levels.}
\label{fig:BER_accuracy-recovery-tradeoff}
\end{figure}
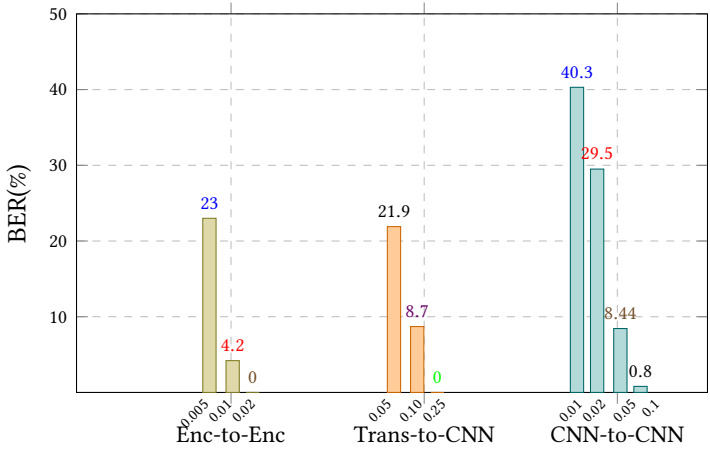

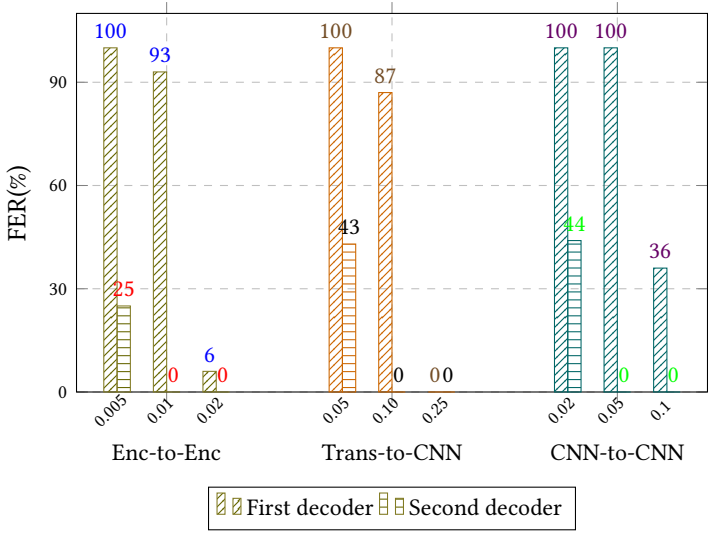
\begin{figure}[htbp]
\centering
\makebox[\columnwidth][c]{%
\begin{minipage}{0.6\columnwidth}
\centering
\begin{tikzpicture}[trim axis left]
\begin{axis}[
    ybar,
    bar width=5pt,
    width=\linewidth,
    height=0.6\linewidth,
    scale only axis,
    xmin=0.6, xmax=3.4,
    ymin=0, ymax=110,
    ylabel={FER(\%)},
    xtick={1,2,3},
    xticklabels={Enc-to-Enc, Trans-to-CNN, CNN-to-CNN},
    xticklabel style={font=\small, yshift=-12pt},
    ylabel style={font=\scriptsize},
    tick label style={font=\scriptsize},
    nodes near coords,
    every node near coord/.append style={font=\small},
    legend style={
        font=\small,
        at={(0.5,-0.25)},
        anchor=north,
        legend columns=2
    },
    axis lines=box,
    clip=false,
    enlarge x limits=0,
    ymajorgrids=true,
    xmajorgrids=true,
    grid style=dashed,
    ytick={0,30,60,90},
]

\pgfmathsetmacro{\sub}{0.22}
\newcommand{\pairshift}{2.5pt}

\def\AtenOne{100.0}   \def\BtenOne{25.0}
\def\AtenTwo{93.0}    \def\BtenTwo{0.0}
\def\AtenFive{6.0}    \def\BtenFive{0.0}

\def\AtwentyOne{100.0} \def\BtwentyOne{43.0}
\def\AtwentyTwo{87.0}  \def\BtwentyTwo{0.0}
\def\AtwentyFive{0.0}  \def\BtwentyFive{0.0}

\def\AfiftyOne{100.0}  \def\BfiftyOne{44.0}
\def\AfiftyTwo{100.0}  \def\BfiftyTwo{0.0}
\def\AfiftyFive{36.0}  \def\BfiftyFive{0.0}

\addplot+[
    draw=olive!80!black,
    fill=olive!30,
    pattern=north east lines,
    pattern color=olive!80!black,
    bar shift=-\pairshift
]
coordinates {
    (1-\sub,\AtenOne)
    (1,\AtenTwo)
    (1+\sub,\AtenFive)
};

\addplot+[
    draw=olive!80!black,
    fill=olive!10,
    pattern=horizontal lines,
    pattern color=olive!80!black,
    bar shift=\pairshift
]
coordinates {
    (1-\sub,\BtenOne)
    (1,\BtenTwo)
    (1+\sub,\BtenFive)
};

\addplot+[
    draw=orange!80!black,
    fill=orange!40,
    pattern=north east lines,
    pattern color=orange!80!black,
    bar shift=-\pairshift
]
coordinates {
    (2-\sub,\AtwentyOne)
    (2,\AtwentyTwo)
    (2+\sub,\AtwentyFive)
};

\addplot+[
    draw=orange!80!black,
    fill=orange!15,
    pattern=horizontal lines,
    pattern color=orange!80!black,
    bar shift=\pairshift
]
coordinates {
    (2-\sub,\BtwentyOne)
    (2,\BtwentyTwo)
    (2+\sub,\BtwentyFive)
};

\addplot+[
    draw=teal!80!black,
    fill=teal!30,
    pattern=north east lines,
    pattern color=teal!80!black,
    bar shift=-\pairshift
]
coordinates {
    (3-\sub,\AfiftyOne)
    (3,\AfiftyTwo)
    (3+\sub,\AfiftyFive)
};

\addplot+[
    draw=teal!80!black,
    fill=teal!10,
    pattern=horizontal lines,
    pattern color=teal!80!black,
    bar shift=\pairshift
]
coordinates {
    (3-\sub,\BfiftyOne)
    (3,\BfiftyTwo)
    (3+\sub,\BfiftyFive)
};

\addlegendentry{First decoder}
\addlegendentry{Second decoder}

\node[font=\scriptsize, anchor=north, rotate=45, yshift=-4pt]
at (axis cs:1-1.5*\sub,0) {0.005};
\node[font=\scriptsize, anchor=north, rotate=45, yshift=-4pt]
at (axis cs:1-0.5*\sub,0) {0.01};
\node[font=\scriptsize, anchor=north, rotate=45, yshift=-4pt]
at (axis cs:1+0.5*\sub,0) {0.02};

\node[font=\scriptsize, anchor=north, rotate=45, yshift=-4pt]
at (axis cs:2-1.5*\sub,0) {0.05};
\node[font=\scriptsize, anchor=north, rotate=45, yshift=-4pt]
at (axis cs:2-0.5*\sub,0) {0.10};
\node[font=\scriptsize, anchor=north, rotate=45, yshift=-4pt]
at (axis cs:2+0.5*\sub,0) {0.25};

\node[font=\scriptsize, anchor=north, rotate=45, yshift=-4pt]
at (axis cs:3-1.5*\sub,0) {0.02};
\node[font=\scriptsize, anchor=north, rotate=45, yshift=-4pt]
at (axis cs:3-0.5*\sub,0) {0.05};
\node[font=\scriptsize, anchor=north, rotate=45, yshift=-4pt]
at (axis cs:3+0.5*\sub,0) {0.1};

\end{axis}
\end{tikzpicture}
\end{minipage}%
}

\caption{Final FER after second-stage recovery across different PUF lengths and perturbation levels.}
\label{fig:FER_accuracy-recovery-tradeoff}
\vspace{-3mm}
\end{figure}

The first stage serves as the primary mechanism for extracting the hardware signature from the student model. As illustrated in Figure~\ref{fig:BER_accuracy-recovery-tradeoff}, the neural decoder demonstrates a significant capacity to recover a large fraction of the fingerprint bits, with BER decreasing as the perturbation strength $\epsilon$ increases. However, the system-level reliability remains poor at low $\epsilon$ values, as shown in Fig.~\ref{fig:FER_accuracy-recovery-tradeoff}. For example, in the \textbf{Enc-to-Enc (CIFAR-10)} setup at $\epsilon=0.005$, the decoder achieves a BER of 23\%, yet the resulting FER is 100\%. This indicates that while the first stage successfully extracts the majority of the signature, the remaining bit errors are sufficient to prevent correct device authentication. We observe a similar trend in the \textbf{CNN-to-CNN (CIFAR-50)} architecture; at $\epsilon=0.02$, a $29.5\%$ initial BER results in a 100\% FER, as every recovery attempt contains at least one bit-flip error. These results highlight a critical gap: the first stage provides a strong foundation for bit recovery but is insufficient for reliable identification on its own. This discrepancy—where bit-level success does not yet translate to error-free authentication—directly motivates the necessity of the second-stage refinement to achieve the $0\%$ FER required for secure hardware-rooted identification.

\subsubsection{Stage-II: Hamming Refinement and Attribution}

The second stage acts as a closest match search to finalize the identification process. By calculating the difference between the recovered bits and the stored keys, the system identifies the enrolled device that is closest to the noisy estimate.

The efficacy of this refinement is evident across all evaluated paradigms, as the Hamming-distance search bridges the gap between noisy neural estimations and deterministic hardware identities. In the Enc-to-Enc (CIFAR-10) workflow at a minimal strength of $\epsilon=0.005$, Stage-II refinement successfully reduces the initial $100\%$ FER down to $25.0\%$, achieving a perfect $0.0\%$ FER once the perturbation is increased to $\epsilon=0.010$. A similar corrective trend is observed in the Trans-to-CNN (CIFAR-20) architecture, where a raw Stage-I FER of $100\%$ at $\epsilon=0.05$ is drastically suppressed to $43\%$ through refinement, eventually achieving perfect recovery 
at $\epsilon=0.10$. Furthermore, the \textbf{CNN-to-CNN (CIFAR-50)} system demonstrates the robustness of the Hamming-distance decoder; despite a failure in Stage-I recovery at $\epsilon=0.02$, Stage-II refinement manages to recover nearly half of the identities ($44\%$ final FER), before achieving reliable identification
($0\%$ FER) at strengths of $\epsilon \geq 0.05$.

\subsection{PUF Length Selection and Identity Scalability}
\label{subsec:bit_compression}

\begin{figure}[t]
\centering
\makebox[\columnwidth][c]{%
\begin{minipage}{0.8\columnwidth}
\centering
\begin{tikzpicture}
\pgfplotsset{compat=1.18}

\begin{axis}[
    name=success,
    width=0.95\linewidth,
    height=4.4cm,
    xmin=0.48, xmax=1.005,
    ymin=45, ymax=100,
    ylabel={Success Rate (\%)},
    ylabel style={font=\small},
    xtick={0.5,0.6,0.7,0.8,0.9,1.0},
    xticklabels={},
    ytick={50,60,70,80,90,100},
    tick label style={font=\scriptsize},
    grid=major,
    grid style={gray!20},
    axis lines=box,
    axis line style={black},
    clip=false
]

\path[fill=gray!20, fill opacity=0.25]
(axis cs:0.95,45) rectangle (axis cs:1.0,100);

\draw[densely dashed, gray!70]
(axis cs:0.95,45) -- (axis cs:0.95,100);

\addplot[
    teal!80!black,
    thick,
    solid,
    mark=*,
    mark size=1.6pt,
    forget plot
] coordinates {
    (0.500,65.20)
    (0.609,77.20)
    (0.700,85.40)
    (0.781,88.40)
    (0.906,91.60)
    (0.922,94.80)
    (0.938,93.60)
    (0.953,91.20)
    (0.969,95.20)
    (0.984,93.20)
    (1.000,98.00)
};

\addplot[
    violet!80!black,
    thick,
    solid,
    mark=square*,
    mark size=1.6pt,
    forget plot
] coordinates {
    (0.500,89.20)
    (0.602,93.20)
    (0.703,95.60)
    (0.805,96.30)
    (0.938,94.40)
    (0.953,98.40)
    (0.969,98.40)
    (0.984,99.50)
    (1.000,99.50)
};

\addplot[
    orange!85!black,
    thick,
    solid,
    mark=triangle*,
    mark size=1.6pt,
    forget plot
] coordinates {
    (0.500,72.00)
    (0.602,83.60)
    (0.703,88.80)
    (0.801,90.00)
    (0.902,96.40)
    (0.945,90.40)
    (0.953,95.60)
    (0.961,94.80)
    (0.969,94.40)
    (0.977,95.60)
    (0.984,96.40)
    (0.992,98.00)
    (1.000,96.00)
};

\node[
    font=\scriptsize,
    anchor=north west,
    text=black
]
at (axis description cs:0.03,0.96)
{(a) Success rate};

\node[
    font=\scriptsize,
    align=center,
    text=gray!40!black
]
at (axis cs:0.58,56.5)
{Shorter keys:\\lower margin};

\node[
    font=\scriptsize,
    align=center,
    text=black,
    anchor=south
]
at (axis description cs:0.90,1.02)
{Recommended region};

\end{axis}

\begin{axis}[
    name=successzoom,
    at={(success.south east)},
    anchor=south east,
    xshift=-0.15cm,
    yshift=0.35cm,
    width=0.48\linewidth,
    height=2.35cm,
    xmin=0.95, xmax=1.001,
    ymin=90, ymax=100,
    xtick={0.95,0.97,0.99,1.00},
    ytick={90,95,100},
    tick label style={font=\scriptsize},
    axis lines=box,
    axis line style={black},
    grid=major,
    grid style={gray!15},
    axis background/.style={fill=white},
    clip=false,
    title={\scriptsize Zoomed success region},
    title style={font=\scriptsize, yshift=-1pt}
]

\addplot[
    teal!80!black,
    thick,
    solid,
    mark=*,
    mark size=1.5pt
] coordinates {
    (0.953,91.20)
    (0.969,95.20)
    (0.984,93.20)
    (1.000,98.00)
};

\addplot[
    violet!80!black,
    thick,
    solid,
    mark=square*,
    mark size=1.5pt
] coordinates {
    (0.953,98.40)
    (0.969,98.40)
    (0.984,99.50)
    (1.000,99.50)
};

\addplot[
    orange!85!black,
    thick,
    solid,
    mark=triangle*,
    mark size=1.5pt
] coordinates {
    (0.953,95.60)
    (0.961,94.80)
    (0.969,94.40)
    (0.977,95.60)
    (0.984,96.40)
    (0.992,98.00)
    (1.000,96.00)
};

\end{axis}

\begin{axis}[
    name=ber,
    at={(success.south west)},
    anchor=north west,
    yshift=-0.45cm,
    width=0.95\linewidth,
    height=3.4cm,
    xmin=0.48, xmax=1.005,
    ymin=0, ymax=25,
    xlabel={Normalized PUF Length $n/d$},
    xlabel style={font=\small, yshift=1pt},
    ylabel={Avg. BER (\%)},
    ylabel style={font=\small},
    xtick={0.5,0.6,0.7,0.8,0.9,1.0},
    ytick={0,5,10,15,20,25},
    tick label style={font=\scriptsize},
    grid=major,
    grid style={gray!20},
    axis lines=box,
    axis line style={black},
    clip=false,
    legend style={
        at={(0.5,-0.52)},
        anchor=north,
        draw=none,
        font=\scriptsize,
        legend columns=3,
        column sep=10pt
    },
    legend cell align={left}
]

\path[fill=gray!20, fill opacity=0.25]
(axis cs:0.95,0) rectangle (axis cs:1.0,25);

\draw[densely dashed, gray!70]
(axis cs:0.95,0) -- (axis cs:0.95,25);

\addlegendimage{only marks, mark=*, mark size=2.2pt, teal!80!black}
\addlegendentry{$d=64$}

\addlegendimage{only marks, mark=square*, mark size=2.2pt, violet!80!black}
\addlegendentry{$d=128$}

\addlegendimage{only marks, mark=triangle*, mark size=2.2pt, orange!85!black}
\addlegendentry{$d=256$}

\addplot[
    teal!80!black,
    thick,
    solid,
    mark=*,
    mark size=1.6pt,
    forget plot
] coordinates {
    (0.500,20.71)
    (0.609,18.09)
    (0.700,15.75)
    (0.781,13.81)
    (0.906,13.57)
    (0.922,10.07)
    (0.938,10.50)
    (0.953,14.11)
    (0.969,12.01)
    (0.984,12.50)
    (1.000,10.96)
};

\addplot[
    violet!80!black,
    thick,
    solid,
    mark=square*,
    mark size=1.6pt,
    forget plot
] coordinates {
    (0.500,14.91)
    (0.602,14.09)
    (0.703,10.88)
    (0.805,9.18)
    (0.938,10.32)
    (0.953,8.19)
    (0.969,7.02)
    (0.984,6.67)
    (1.000,7.78)
};

\addplot[
    orange!85!black,
    thick,
    solid,
    mark=triangle*,
    mark size=1.6pt,
    forget plot
] coordinates {
    (0.500,21.19)
    (0.602,15.94)
    (0.703,12.56)
    (0.801,11.66)
    (0.902,10.97)
    (0.945,11.60)
    (0.953,10.71)
    (0.961,10.19)
    (0.969,9.13)
    (0.977,10.67)
    (0.984,10.32)
    (0.992,8.44)
    (1.000,9.73)
};

\node[
    font=\scriptsize,
    anchor=north west,
    text=black
]
at (axis description cs:0.03,0.93)
{(b) Average BER};

\end{axis}

\end{tikzpicture}
\end{minipage}%
}

\vspace{0.15cm}
\caption{PUF-length tradeoff for $d=64$, $d=128$, and $d=256$ under normalized PUF lengths $n/d$. The top panel reports final success rate after Stage-II Hamming matching, and the bottom panel reports average BER after the Stage-I decoder.}
\label{fig:puf_length_tradeoff_all}
\end{figure}
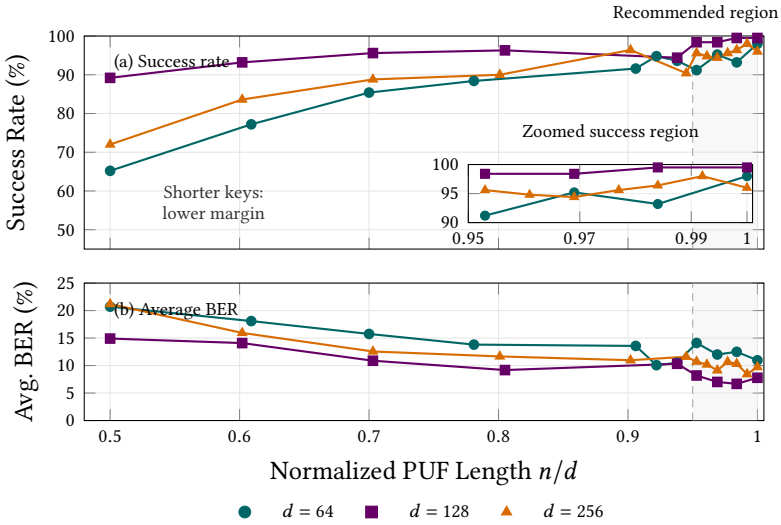

Fig.~\ref{fig:puf_length_tradeoff_all} evaluates how the PUF length $n$ should be selected relative to the logit/latent dimension $d$ when $n \leq d$. For each fixed dimension $d$, we sweep different PUF lengths and report the normalized ratio $n/d$. Within each dimension group, the teacher--student architecture, dataset, enrolled-device setting, and perturbation strength $\epsilon$ are kept fixed, while $\epsilon$ is calibrated separately for each $d$ to keep the PUF signal recoverable under the corresponding dimensional constraint. Each configuration is evaluated using multiple independently generated projection matrices $U$, and the curves report the mean success rate after Stage-II Hamming matching and the mean BER after the Stage-I decoder. Although the standard-deviation bars are omitted from the figure for readability, the measured deviations show that recovery generally becomes more stable when $n$ is close to $d$. For example, when $d=128$, the BER standard deviation is $5.03\%$ at $n/d=0.938$, but decreases to $1.54\%$ at $n/d=0.953$ and $0.09\%$ at $n/d=1.0$. 

The results show that shorter PUF signatures tend to produce lower success rate (higher FER) and higher BER, likely because they provide weaker inter-device separation and make Stage-II matching more sensitive to Stage-I bit errors. As $n/d$ increases, the enrolled keys become more discriminative, improving recovery reliability. Based on these observations, we identify $0.95 \leq n/d \leq 1.0$ as the recommended practical operating region, which provides high success rate, controlled BER, and reduced projection-dependent variation across the evaluated dimensions. Therefore, in practical deployment, selecting $n$ close to the logit/latent dimension provides a stable tradeoff between device-code discriminability and recovery robustness.

\rev{We then evaluate the high-capacity regime where the PUF length exceeds
the output dimension. For CIFAR-10, the output space contains 10 classes,
which limits the number of directly embedded identity bits. To scale beyond
this limit, we use Multi-Level Encoding, where each output dimension carries
multiple PUF bits through discrete perturbation levels. This allows the
identity length to increase from 10 bits to 20 and 30 bits in the CNN-to-CNN
distillation setting. The results show a clear trade-off between fingerprint capacity, recovery
reliability, and model utility. The baseline CNN student accuracy is $82.4\%$
when $\epsilon=0$. With a 10-bit key, the student accuracy remains
$81.67 \pm 0.55\%$, while $\epsilon=0.05$ is sufficient to achieve
$0\%$ FER. For a 20-bit key, stronger perturbation is required:
$\epsilon=0.1$ still results in $32\%$ FER, whereas $\epsilon=0.2$
achieves reliable recovery. For the 30-bit setting, $\epsilon=0.4$ is
required to reach $0\%$ FER, and a larger accuracy variation is observed,
with a standard deviation of up to $16.54\%$. These results indicate that
packing more PUF bits into each logit increases identity capacity but reduces
the available utility margin, requiring careful selection of the perturbation
strength.}


\begin{figure}[htbp]
\centering
\begin{tikzpicture}
\begin{axis}[
    ybar,
    bar width=5pt,
    width=0.7\columnwidth,
    height=0.45\linewidth,
    xmin=0, xmax=3.7,
    ymin=0, ymax=110,
    ylabel={FER/Acc.(\%)},
    xtick={0.3, 2, 3.7},
    xticklabels={10-bit, 20-bit, 30-bit},
    xticklabel style={font=\small, yshift=-5pt},
    ylabel style={font=\scriptsize},
    tick label style={font=\scriptsize},
    every node near coord/.append style={font=\small},
    legend style={
        font=\small,
        at={(0.5,-0.18)},
        anchor=north,
        legend columns=2
    },
    axis lines=box,
    clip=false,
    enlarge x limits=0.2,
    ymajorgrids=true,
    xmajorgrids=true,
    grid style=dashed,
    ytick={20,40,60,80,100},
    extra x tick style={grid style=dashed},
]

\pgfmathsetmacro{\shift}{0.06}

\addplot+[
    draw=blue,
    fill=blue!30,
    error bars/.cd,
    y dir=both,
    y explicit
] coordinates {
    (1 - 0.5*\shift, 81.67) +- (0, 0.55)
};

\node[
    font=\scriptsize,
    anchor=east,
    xshift=-32pt,
    yshift=4pt
] at (axis cs:1 - 0.5*\shift, 81.67)
{81.67 $\pm$ 0.55};

\addplot+[
    draw=red,
    fill=red!30
] coordinates {
    (1 - 0.5*\shift, 15.0)
};

\node[
    font=\scriptsize,
    anchor=south,
    xshift=-32pt
] at (axis cs:1 - 0.5*\shift, 15.0)
{15};

\addplot+[
    draw=blue,
    fill=blue!30,
    error bars/.cd,
    y dir=both,
    y explicit
] coordinates {
    (1 - 0.1*\shift, 81.16) +- (0, 0.56)
};

\node[
    font=\scriptsize,
    anchor=east,
    xshift=2pt,
    yshift=4pt
] at (axis cs:1 - 0.1*\shift, 86.67)
{81.67 $\pm$ 0.56};

\addplot+[
    draw=red,
    fill=red!30
] coordinates {
    (1 - 0.3*\shift, 0.0)
};

\node[
    font=\scriptsize,
    anchor=south,
    xshift=-17pt
] at (axis cs:1 - 0.3*\shift, 0.0)
{0};

\node[
    font=\scriptsize,
    anchor=north east,
    rotate=45
] at (axis cs:0.1, 2)
{0.02};

\node[
    font=\scriptsize,
    anchor=north east,
    rotate=45
] at (axis cs:0.5, 2)
{0.05};

\addplot+[
    draw=blue,
    fill=blue!30,
    error bars/.cd,
    y dir=both,
    y explicit
] coordinates {
    (2 - 0.5*\shift, 65.99) +- (0, 8.31)
};

\node[
    font=\scriptsize,
    anchor=east,
    xshift=-2pt,
    yshift=15pt
] at (axis cs:2 - 0.5*\shift, 65.99)
{65.99 $\pm$ 8.31};

\addplot+[
    draw=red,
    fill=red!30
] coordinates {
    (2 - 0.3*\shift, 32.0)
};

\node[
    font=\scriptsize,
    anchor=south,
    xshift=-4pt
] at (axis cs:2 - 0.3*\shift, 32.0)
{32};

\addplot+[
    draw=blue,
    fill=blue!30,
    error bars/.cd,
    y dir=both,
    y explicit
] coordinates {
    (2 - 0.1*\shift, 70.22) +- (0, 10.19)
};

\node[
    font=\scriptsize,
    anchor=east,
    xshift=25pt,
    yshift=20pt
] at (axis cs:2 - 0.1*\shift, 70.22)
{70.22 $\pm$ 10.19};

\addplot+[
    draw=red,
    fill=red!30
] coordinates {
    (2 + 0.3*\shift, 0.0)
};

\node[
    font=\scriptsize,
    anchor=south,
    xshift=10pt
] at (axis cs:2 + 0.3*\shift, 0.0)
{0};

\node[
    font=\scriptsize,
    anchor=north east,
    rotate=45
] at (axis cs:1.8, 2)
{0.1};

\node[
    font=\scriptsize,
    anchor=north east,
    rotate=45
] at (axis cs:2.2, 2)
{0.2};

\addplot+[
    draw=blue,
    fill=blue!30,
    error bars/.cd,
    y dir=both,
    y explicit
] coordinates {
    (3 - 0.5*\shift, 83.22) +- (0, 15.65)
};

\node[
    font=\scriptsize,
    anchor=east,
    xshift=20pt,
    yshift=25pt
] at (axis cs:3 - 0.5*\shift, 83.22)
{83.22 $\pm$ 15.65};

\addplot+[
    draw=red,
    fill=red!30
] coordinates {
    (3 - 0.3*\shift, 57.0)
};

\node[
    font=\scriptsize,
    anchor=south,
    xshift=25pt
] at (axis cs:3 - 0.3*\shift, 57.0)
{57};

\addplot+[
    draw=blue,
    fill=blue!30,
    error bars/.cd,
    y dir=both,
    y explicit
] coordinates {
    (3 - 0.1*\shift, 82.82) +- (0, 16.54)
};

\node[
    font=\scriptsize,
    anchor=east,
    xshift=61pt,
    yshift=25pt
] at (axis cs:3 - 0.1*\shift, 82.82)
{82.82 $\pm$ 16.54};

\addplot+[
    draw=red,
    fill=red!30
] coordinates {
    (3 + 0.3*\shift, 0.0)
};

\node[
    font=\scriptsize,
    anchor=south,
    xshift=37pt
] at (axis cs:3 + 0.3*\shift, 0.0)
{0};

\node[
    font=\scriptsize,
    anchor=north east,
    rotate=45
] at (axis cs:3.5, 2)
{0.3};

\node[
    font=\scriptsize,
    anchor=north east,
    rotate=45
] at (axis cs:3.9, 2)
{0.4};

\legend{Student Accuracy, PUF FER}

\end{axis}
\end{tikzpicture}

\caption{Trade-off between student classification accuracy (CIFAR-10) and Final FER with CNN-to-CNN architecture for Multi-Level Encoding scheme.}
\label{fig:compressed-accuracy-recovery-tradeoff}
\end{figure}
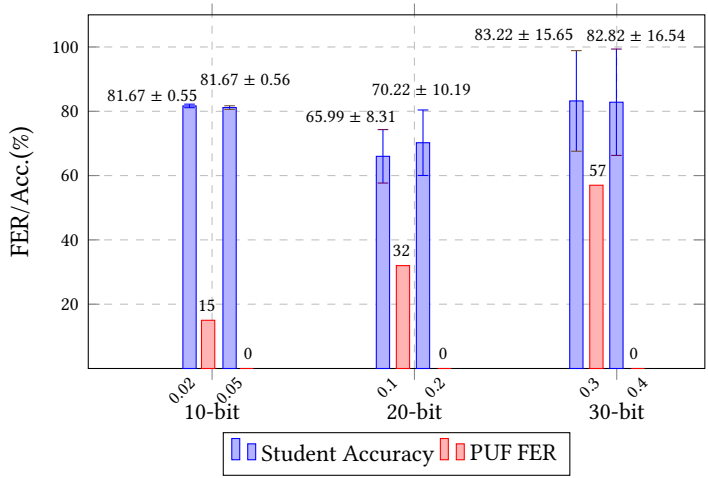

\begin{figure*}[t]
\centering

\begin{center}
\pgfplotslegendfromname{shared_legend_name}
\end{center}
\vspace{2pt}

\begin{minipage}{0.33\textwidth}
\centering
\begin{tikzpicture}
\begin{axis}[
    width=\textwidth,
    height=4.5cm,
    axis lines=left,
    grid=both,
    xmin=0, xmax=50,
    ymin=0, ymax=40,
    ylabel={Raw BER (\%)},
    xlabel={Training Epochs},
    xtick={0,25,50},
    legend to name=shared_legend_name,
    legend style={
        font=\small,
        draw=none,
        legend columns=3,
        column sep=4pt
    }
]

\addplot[green!60!black, mark=square*, thick]
coordinates {(0,1.6) (25,2) (50,1.9)};
\addlegendentry{Downstream FT}

\addplot[orange, mark=triangle*, thick]
coordinates {(0,4.7) (25,11) (50,17.5)};
\addlegendentry{Diff Dataset FT}

\addlegendimage{blue, mark=*, thick}
\addlegendentry{Latent Noise}

\addlegendimage{red, dashed, mark=triangle*, thick}
\addlegendentry{SSIM}

\addlegendimage{violet!80!black, dashed, mark=diamond*, thick}
\addlegendentry{$D_W(t)$}

\end{axis}
\end{tikzpicture}
\end{minipage}%
\hfill%
\begin{minipage}{0.33\textwidth}
\centering
\begin{tikzpicture}
\begin{axis}[
    width=\textwidth,
    height=4.5cm,
    axis lines=left,
    grid=both,
    xmin=0, xmax=50,
    ymin=0, ymax=40,
    ylabel={Final BER (\%)},
    xlabel={Training Epochs},
    xtick={0,25,50}
]

\addplot[green!60!black, mark=square*, thick]
coordinates {(0,0) (25,0) (50,0)};

\addplot[orange, mark=triangle*, thick]
coordinates {(0,0) (25,0) (50,5.6)};

\end{axis}
\end{tikzpicture}
\end{minipage}%
\hfill%
\begin{minipage}{0.33\textwidth}
\centering
\begin{tikzpicture}

\begin{axis}[
    width=\textwidth,
    height=4.5cm,
    axis y line*=left,
    axis x line=bottom,
    grid=both,
    xmin=0, xmax=50,
    ymin=0, ymax=120,
    ytick={0,20,40,60,80,100,120},
    ylabel={Success Rate (\%)},
    xlabel={Training Epochs},
    xtick={0,25,50}
]

\addplot[green!60!black, mark=square*, thick]
coordinates {(0,100) (25,100) (50,100)};

\addplot[orange, mark=triangle*, thick]
coordinates {(0,100) (25,100) (50,87.5)};

\end{axis}

\begin{axis}[
    width=\textwidth,
    height=4.5cm,
    axis y line*=right,
    axis x line=none,
    xmin=0, xmax=50,
    ymin=0, ymax=0.7,
    ytick={0,0.1,0.2,0.3,0.4,0.5,0.6,0.7},
    ylabel={$D_W(t)$},
    ylabel style={violet!80!black},
    yticklabel style={violet!80!black}
]

\addplot[violet!80!black, dashed, mark=diamond*, thick]
coordinates {
    (0,0)
    (1,0.055)
    (5,0.117)
    (10,0.185)
    (20,0.311)
    (30,0.426)
    (50,0.626)
};

\end{axis}
\end{tikzpicture}
\end{minipage}

\vspace{0.1cm}

\begin{minipage}{0.33\textwidth}
\centering
\begin{tikzpicture}
\begin{axis}[
    width=\textwidth,
    height=4.5cm,
    axis lines=left,
    grid=both,
    xmin=0, xmax=0.15,
    ymin=0, ymax=40,
    ylabel={Raw BER (\%)},
    xlabel={$\sigma$}
]

\addplot[blue, mark=*, thick]
coordinates {(0,5.7) (0.1,21) (0.15,28)};

\end{axis}
\end{tikzpicture}
\end{minipage}%
\hfill%
\begin{minipage}{0.33\textwidth}
\centering
\begin{tikzpicture}
\begin{axis}[
    width=\textwidth,
    height=4.5cm,
    axis lines=left,
    grid=both,
    xmin=0, xmax=0.15,
    ymin=0, ymax=40,
    ylabel={Final BER (\%)},
    xlabel={$\sigma$}
]

\addplot[blue, mark=*, thick]
coordinates {(0,0) (0.1,0) (0.15,15)};

\end{axis}
\end{tikzpicture}
\end{minipage}%
\hfill%
\begin{minipage}{0.33\textwidth}
\centering
\begin{tikzpicture}

\begin{axis}[
    width=\textwidth,
    height=4.5cm,
    axis y line*=left,
    axis x line=bottom,
    grid=both,
    xmin=0, xmax=0.15,
    ymin=0, ymax=120,
    ytick={0,20,40,60,80,100,120},
    ylabel={Success Rate (\%)},
    xlabel={$\sigma$}
]

\addplot[blue, mark=*, thick]
coordinates {(0,100) (0.1,100) (0.15,69)};

\end{axis}

\begin{axis}[
    width=\textwidth,
    height=4.5cm,
    axis y line*=right,
    axis x line=none,
    xmin=0, xmax=0.15,
    ymin=0, ymax=120,
    ytick={0,20,40,60,80,100,120},
    ylabel={SSIM (\%)},
    ylabel style={red},
    yticklabel style={red}
]

\addplot[red, dashed, mark=triangle*, thick]
coordinates {(0,60.9) (0.1,57.7) (0.15,54)};

\end{axis}

\end{tikzpicture}
\end{minipage}

\vspace{0.05cm}

\caption{
Robustness analysis under fine-tuning (top row) and noise injection
(bottom row). The top-right panel additionally reports normalized weight
displacement $D_W(t)$ during post-training.
}
\label{fig:2by3}

\end{figure*}
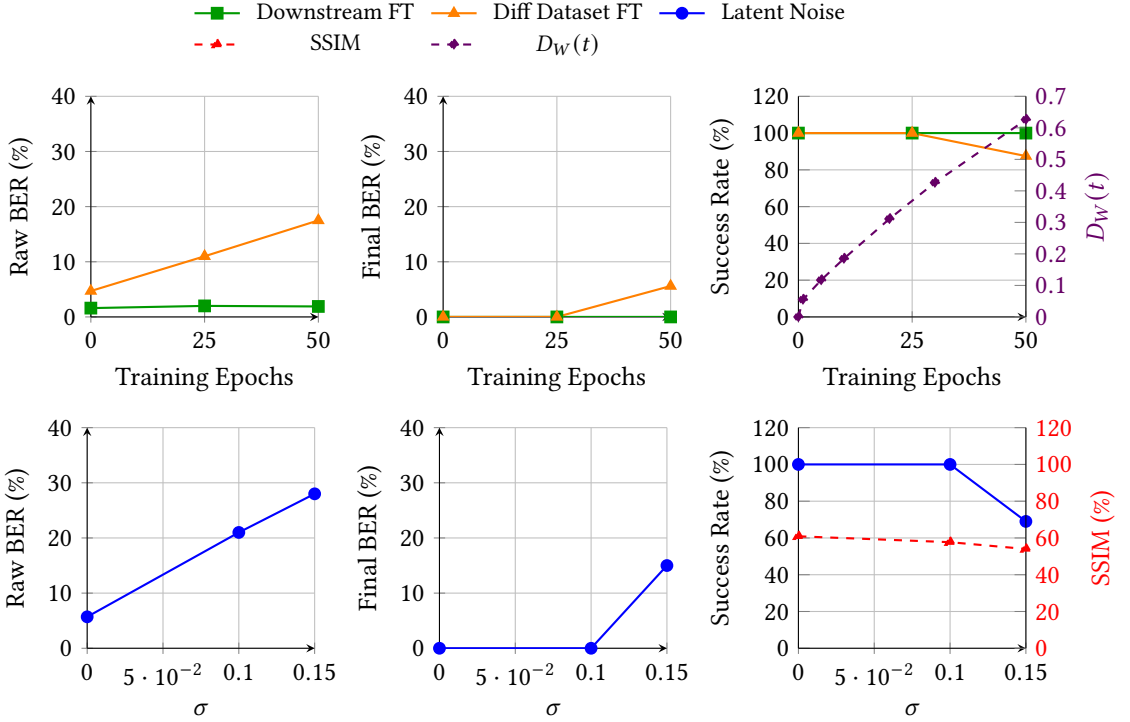

\subsection{Robustness Against Post-Distillation Model Modifications}
\label{sec:robustness}

Once deployed, a distilled student model may undergo further modifications due to routine optimization or deliberate attempts to remove the embedded hardware identity. To evaluate the robustness of the proposed PUF-based fingerprinting framework, we consider three representative post-distillation transformations: in-domain fine-tuning, cross-domain adaptation, and latent-space stochastic corruption.

For each device, we first perform PUF recovery on the trained student model prior to any modification. We then apply each transformation independently and repeat the recovery process using the same decoder. Performance is evaluated using both bit-level and device-level metrics, including raw BER (before first-stage recovery), final BER (after two-stage recovery), and recovery success rate. We additionally report Structural Similarity Index Measure (SSIM) to quantify model utility. \rev{To quantify the extent of parameter modification during post-training, we additionally report the normalized weight displacement
$D_W(t)=\frac{\|\mathbf{W}_t-\mathbf{W}_0\|_2}{\|\mathbf{W}_0\|_2}$,
where $\mathbf{W}_0$ denotes the fingerprinted student parameters before post-training and $\mathbf{W}_t$ denotes the parameters after $t$ additional training epochs.}

\subsubsection{Fine-Tuning Attack: In-Domain and Cross-Domain}

We consider a realistic post-distillation scenario in which the released student model consists of an encoder-only architecture. Since the encoder does not directly produce pixel-space outputs, an adversary may augment the model with a lightweight decoder to enable reconstruction-based optimization.
Under this setting, the attacker performs additional fine-tuning using a standard reconstruction objective, defined as MSE between the input image $\mathbf{x}$ and its reconstruction $\hat{\mathbf{x}}$:
$\mathcal{L}_{\text{MSE}} = \frac{1}{N} \sum_{i=1}^{N} \left\| \hat{\mathbf{x}}_i - \mathbf{x}_i \right\|_2^2$,
where $\hat{\mathbf{x}}_i = D(E(\mathbf{x}_i))$, with $E(\cdot)$ denoting the distilled encoder and $D(\cdot)$ representing the attacker-introduced decoder.

Importantly, the fine-tuning process only aims to improve reconstruction and does not try to keep the embedded hardware identity. This tests whether normal retraining can wash out the fingerprint. As shown in Fig.~\ref{fig:2by3} (top row), \textbf{in-domain fine-tuning (Downstream FT) with CIFAR-10 has negligible impact on fingerprint recovery}. Both raw and final BER remain close to zero, and the recovery success rate stays at 100\% even after 50 additional epochs-a duration matching the original training budget of the teacher model. This suggests that fingerprints embedded in task-aligned representations are inherently stable under standard optimization. 

In contrast, \textbf{cross-domain fine-tuning (Diff Dataset FT) with dataset STL-10 introduces limited degradation}. While recovery remains stable for moderate training (up to 25 epochs), extended fine-tuning (50 epochs) increases the final BER to approximately 6\% and and reduces the success rate to approximately 90\%. However, this attack duration equals the original 50-epoch training budget of the teacher model and significantly exceeds the 30-epoch distillation process required to generate the student. From an economic perspective, a 50-epoch ``cleaning" phase effectively doubles the total IP development cost. For an adversary whose primary motivation is the rapid and low-cost acquisition of proprietary model functionality, such a requirement imposes a substantial additional training cost, rendering the theft nearly as expensive as legitimate model development.  These results suggest that while cross-domain adaptation can partially weaken the embedded signature, it does so inefficiently, requiring prolonged retraining while still preserving high traceability.

\rev{Nearly all floating-point weights change numerically after only a small
number of post-training updates, indicating that the fraction of changed
weights alone is not a meaningful robustness threshold. As shown in
Fig.~\ref{fig:2by3} (top-right), the normalized weight displacement increases
progressively with continued training. Despite this growing parameter drift,
Stage-II Hamming matching still preserves correct device attribution in many
evaluated cases.}

\subsubsection{Latent Noise Injection Attack}

We further evaluate the robustness of the embedded fingerprint under direct perturbations applied to the latent representation. Unlike fine-tuning-based attacks, which require additional training, this setting models a low-cost adversarial strategy that modifies the internal feature representation at inference time.

Specifically, given the latent embedding $\mathbf{z}_s$ produced by the student encoder, the attacker injects additive Gaussian noise:
$
\mathbf{z}_s' = \mathbf{z}_s + \boldsymbol{\epsilon}, \quad \boldsymbol{\epsilon} \sim \mathcal{N}(0, \sigma^2 \mathbf{I}),
$
where $\sigma$ controls the perturbation strength. The perturbed latent $\mathbf{z}_s'$ is then used for both reconstruction (via the decoder) and fingerprint recovery.
This attack directly disrupts the structured perturbation introduced by the PUF-based embedding. As shown in Fig.~\ref{fig:2by3} (bottom row), increasing $\sigma$ progressively degrades fingerprint recovery. At $\sigma = 0.15$, the recovery success rate drops to approximately 70\%, with a corresponding increase in final BER.
However, this reduction in traceability is accompanied by a noticeable degradation in model utility. To quantify this effect, we measure the  SSIM\cite{SSIM}.
SSIM captures perceptual structural similarity between the original image $\mathbf{x}$ and its reconstruction $\hat{\mathbf{x}}$. The final SSIM score is obtained by averaging over all local windows across the image. 
As the noise level increases, the SSIM decreases from approximately 60\% to 54\%, indicating a loss in structural fidelity. This demonstrates a clear \textbf{utility–security trade-off}: effective signature suppression requires perturbations that simultaneously impair the model’s functional performance.

Importantly, while the observed SSIM degradation appears moderate in reconstruction tasks, the impact of latent perturbations is significantly more severe in representation learning scenarios such as face recognition. In such settings, even small shifts in the latent space can distort the embedding geometry, leading to substantial errors \cite{fr,fr1} in identity matching or clustering. This highlights that latent noise injection not only weakens traceability but can also critically undermine downstream task reliability.

Across all evaluated scenarios, the proposed PUF-based fingerprinting framework exhibits strong persistence. In-domain optimization does not affect the embedded identity, cross-domain adaptation requires substantial retraining to achieve limited degradation, and direct noise-based attacks incur a clear loss in model utility. These results indicate that removing the hardware-rooted fingerprint is either ineffective under standard training procedures or requires aggressive modifications that degrade model utility.

\section{Conclusion}
\label{sec:conclusion}


This work presents a hardware-rooted fingerprinting framework for  device-level traceability of AI intellectual property. By embedding RO-PUF-derived device signatures into teacher-model logits, the proposed method enables hardware-linked identity inheritance through knowledge distillation. Experimental results show that the fingerprints are recoverable across heterogeneous teacher--student architectures, including CNNs, Vision Transformers, and encoder-based models. The two-stage recovery pipeline, combining neural decoding with Hamming-distance refinement, improves attribution reliability under PUF variation, distillation noise, fine-tuning, and cross-domain adaptation. Overall, the framework provides a scalable mechanism to trace a suspicious student model back to its source hardware.

\bibliographystyle{ACM-Reference-Format}
\bibliography{sample-base}

\end{document}